\documentclass[journal]{vgtc}                     % final (journal style)
\onlineid{2038}

\vgtccategory{Research}

\title{Coherent Visualization of 2D Scalar Field Contour Ensembles With Probabilistic Latent Space Modeling}

\author{%
  Cenyang Wu, Runhao Lin,
  Qinhan Yu, and 
  Liang Zhou*
}

\authorfooter{
  \item
  	Cenyang Wu and Liang Zhou are with the Institute of Medical Technology, Peking University Health Science Center and National Institute of Health Data Science, Peking University.\\
  	E-mails: wuceny@stu.pku.edu.cn, zhoulng@pku.edu.cn.\\
    Liang Zhou is the corresponding author.
        \item
   Runhao Lin is with the National Institute of Health Data Science, Peking University and Beijing Forestry University.
  	E-mail: meruem@bjfu.edu.cn
    \item
  	Qinhan Yu is with the Center for Machine Learning Research, Peking University.
  	E-mail: yuqinhan@stu.pku.edu.cn.
}

\abstract{%
We present a new visualization method for contour ensembles through probabilistic modeling. 
We aim to improve the coherence between different visual representations, such as contour boxplots and density plots for a 2D scalar field ensemble. 
We model each ensemble member with a probabilistic representation in the latent space, i.e., a lower-dimensional representation of spatial data features, of a variational autoencoder (VAE).
Thereafter, efficient data depth computation and uncertainty-aware clustering are supported based on a matrix of pair-wise similarity measurements of members.
We estimate the underlying probability distribution by leveraging the power of VAE to create density plots that align more coherently with member distributions than existing methods.
The effectiveness of our method is evaluated through numerical comparisons with existing techniques, and visualization examples of synthetic and real-world ensemble datasets. 
}


\teaser{
  \centering
  \includegraphics[width=\linewidth]{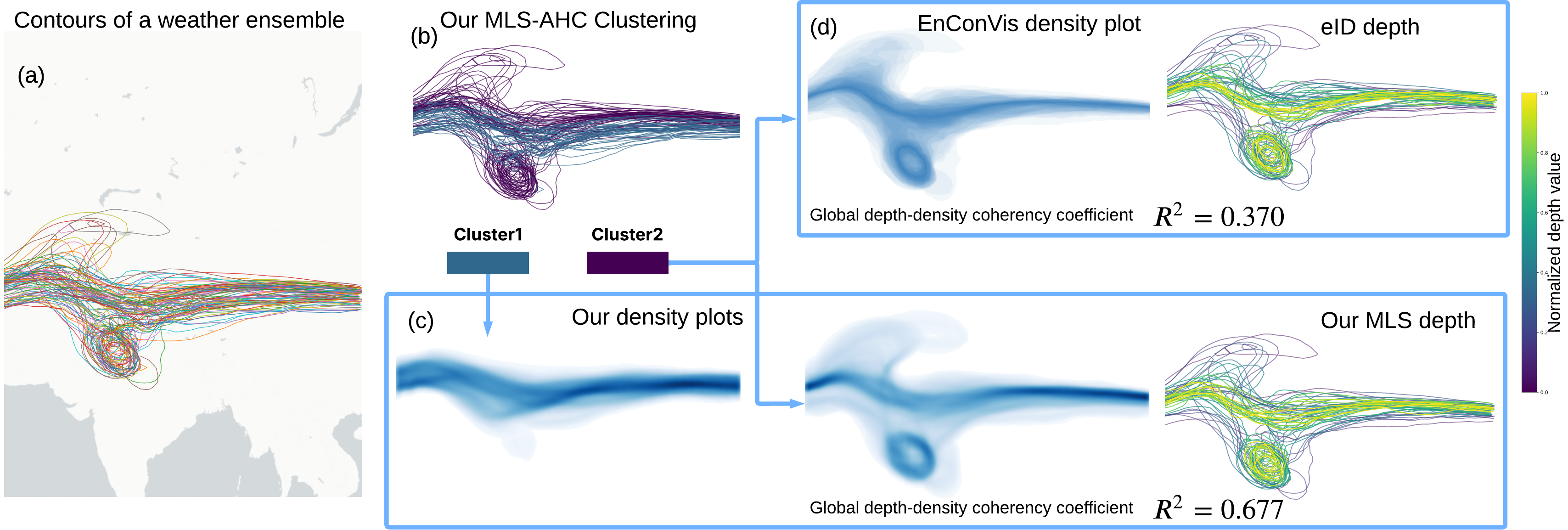}
  \caption{%
 Visualization of an ensemble of a weather simulation with our new probabilistic latent space modeling method. 
 The spaghetti plot of contours from the ensemble is shown in (a).
 Our new uncertainty-aware clustering technique (b) groups the ensembles into two clusters in the latent space. With (c) a coherent processing of the data depth and density plots, our method achieves good depth-density coherence as shown by the depth-density coherency coefficient $R^2=0.677$. For comparison (d), a state-of-the-art density plot method (EnConVis) is less coherent $R^2=0.370$ with a data depth method (eID). 
  }
  \label{fig:teaser}
}

\graphicspath{{figs/}{figures/}{pictures/}{images/}{./}} % where to search for the images

\usepackage{tabu}                      % only used for the table example
\usepackage{booktabs}                  % only used for the table example
\usepackage{lipsum}                    % used to generate placeholder text
\usepackage{mwe}                       % used to generate placeholder figures
\usepackage{ccicons}                   % package to be able to use icons from creative commons

\usepackage{mathptmx}                  % use matching math font
\usepackage{algorithm}
\usepackage{algpseudocode}
\usepackage{amsfonts}
\usepackage{amsmath}

\newcommand{\rev}[1]{{#1}}
\begin{document}

%%%%%%%%%%%%%%%%%%%%%%%%%%%%%%%%%%%%%%%%%%%%%%%%%%%%%%%%%%%%%%%%
%%%%%%%%%%%%%%%%%%%%%% START OF THE PAPER %%%%%%%%%%%%%%%%%%%%%%
%%%%%%%%%%%%%%%%%%%%%%%%%%%%%%%%%%%%%%%%%%%%%%%%%%%%%%%%%%%%%%%%

\firstsection{Introduction}

\maketitle
\label{sec:introduction}
Visualizing spatial features such as contours is essential for the analysis of ensemble datasets that are commonly used for studying the variability and uncertainty of complex systems.
A faithful and uncluttered ensemble visualization is critical for the analysis.

Data depth assigns each data item a centrality score relative to the ensemble distribution, inducing a center-outward ordering from representative to less representative members.
For contour ensembles, this ordering supports robust, nonparametric statistical summaries such as contour boxplots~\cite{whitakerContourBoxplotsMethod2013,chaves-de-plazaInclusionDepthContour2024,wu2025probabilisticinclusiondepthfuzzy}.
These methods offer an accurate statistical visualization of the given ensemble members but do not estimate the probabilistic distribution that they were drawn from.

In contrast, studying the underlying distribution of an ensemble is another common strategy.
Density visualizations approximate the probabilistic density of ensembles ~\cite{Pfaffelmoser2013,kumpfVisualizingConfidenceClusterbased2018,zhangEnConVisUnifiedFramework2023}.
Spatial uncertainty bands characterize the variability of curves or contours using the principal component analysis (PCA)~\cite{ferstlStreamlineVariabilityPlots2016,ferstlVisualAnalysisSpatial2016}.

Branches of techniques in ensemble visualization complement each other by addressing different aspects of the data.
However, one often overlooked issue is that the results from these methods do not align well in the spatial domain.
Evidence of this misalignment can be found in the examples of various studies~\cite{zhangEnConVisUnifiedFramework2023,ferstlStreamlineVariabilityPlots2016} and~\cref{fig:teaser},~\cref{fig:densityEval}.
Such representational inconsistency can complicate interpretation because the visualizations may encode different notions of centrality.

\begin{figure*}[tb] 
    \centering
    \includegraphics[width=\linewidth]{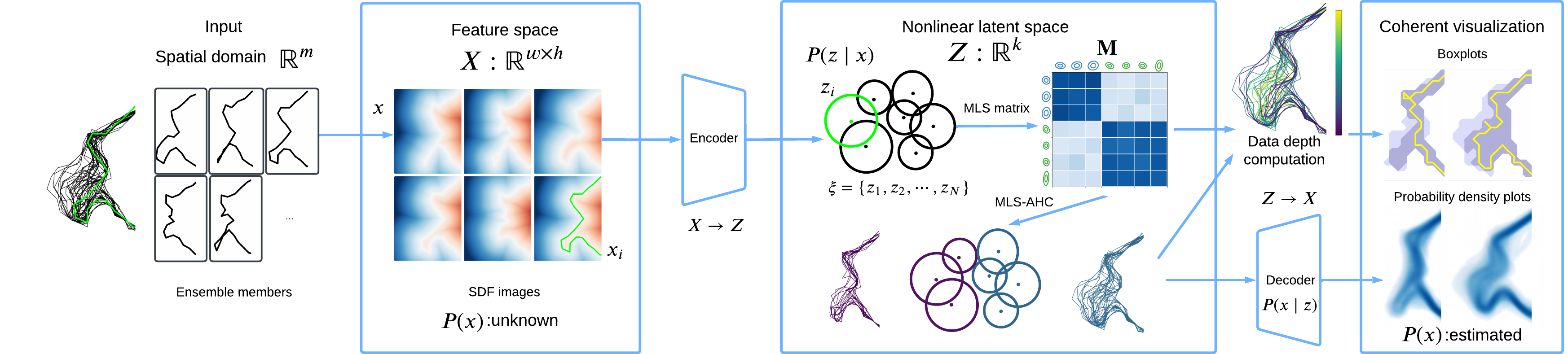}  
    \caption{The workflow of our probabilistic latent space ensemble visualization method. For a contour ensemble, the members are transformed to a feature space and then the nonlinear latent space of VAE as probability distributions (a member is highlighted in green). With a matrix of pairwise
similarity measurements, data depth and clustering are supported in the latent space. By dense sampling the member distributions and transforming
back to the spatial domain, our method generates probability density plots that are coherent with ensemble members. }  
\vspace{-1em}
    \label{fig:workflow}  
\end{figure*}
To bridge this gap, we introduce a probabilistic latent space modeling method for coherent contour ensemble visualization. 
Our overall framework uses the nonlinear latent space of a VAE to model ensemble members. 
Under a probabilistic distribution representation in this latent space, we take a similarity measurement of a pair of ensemble members.
We devise an efficient data depth method for contour boxplot visualization, and an uncertainty-aware clustering technique for multimodal ensemble distributions based on a precomputed matrix.
By sampling around the ensemble members in the latent space, we leverage the generative power of VAE to estimate the probability density and confidence interval bands of the ensemble with high coherence with the given members.
Our method is evaluated with numerical comparisons for the data depth computation, clustering performances, and the coherence of different representation modes compared to state-of-the-art methods.
The usefulness of our method is demonstrated through examples on synthetic and real-world contour ensembles.

An example of a weather simulation ensemble is visualized in~\cref{fig:teaser}.
The ensemble members are visualized in a spaghetti plot (\cref{fig:teaser}(a)) that suffers from occlusion.
Two groups of ensemble members are clustered using our uncertainty-aware clustering technique (\cref{fig:teaser}(b)). 
For each cluster, our method computes the data depth in the latent space (\cref{fig:teaser}(c)-right), and generates coherent density plots (\cref{fig:teaser}(c)-left). 
In comparison, a state-of-the-art density method~\cite{zhangEnConVisUnifiedFramework2023} does not align with the ensemble members indicated by a depth method~\cite{chaves-de-plazaInclusionDepthContour2024} (\cref{fig:teaser}(d)) as well as our method, both visually and numerically.

These are the main contributions of our method.
\begin{itemize}
    \item A probabilistic model in the nonlinear latent space of VAE for contour ensembles that supports various important analysis tasks. 
    \item Efficient data depth computation and uncertainty-aware hierarchical clustering algorithms using a similarity measurement of multidimensional distributions.
    \item A density plot generation method that produces results with good alignment with member distributions.
\end{itemize}

The main benefit of our method is that it provides improved coherence between the data depth representation and density representation compared to existing techniques.
Another benefit is the efficiency and accuracy in computing data depth and ensemble clustering by our method.
We believe that our method can lead to more consistent interpretations across different modes of ensemble visualization.
[We will provide the source code of our method with our paper, if accepted.]

\section{Related Work}
We discuss visualization methods for ensemble feature members, such as contours or curves, and techniques for modeling the probabilistic distributions that generate them.
An overview of the general context of ensemble visualization can be found elsewhere~\cite{Wang2019}.

The spaghetti plot is an intuitive way for ensemble feature visualization as it draws individual ensemble members~\cite{Potter2009,Sanyal2010,Zhang2021}.
However, the visualization easily becomes cluttered when more members are included.
Therefore, a core research question in ensemble visualization is how to reduce clutter while maintaining the faithfulness of the data.

\subsection{Contour Depth Methods}
One strategy is to visualize ensemble members using the data depth.
With a contour band depth definition, order statistics of contours~\cite{whitakerContourBoxplotsMethod2013} and curves~\cite{mirzargarCurveBoxplotGeneralization2014a} are calculated to generate boxplot visualizations for these features in 2D and even 3D~\cite{rajEvaluatingShapeAlignment2016}.
However, the contour band depth does not scale well and is relatively slow.
The epsilon inclusion depth is introduced to improve the computational efficiency of data depth computation~\cite{chaves-de-plazaInclusionDepthContour2024,chaves-de-plazaDepthMultimodalContour2024}.
More recently, a generalization of data depth to fuzzy contours is available with efficient GPU computation to support binary and probabilistic contours of 3D ensembles~\cite{wu2025probabilisticinclusiondepthfuzzy}.

Most data depth techniques assume a unimodal distribution of ensemble members, i.e., all members belong to one cluster. 
This assumption yields false results when the system is actually multimodal~\cite{Pandolfo2023,chaves-de-plazaDepthMultimodalContour2024}.
The multi-modal contour depth method clusters ensemble members by maximizing the average relative depth~\cite{chaves-de-plazaDepthMultimodalContour2024}.

Our new method extends the data depth notion from points to distributions.
We use a similarity matrix for efficient data depth computation, and the matrix also allows for clustering members in multi-modal distributions through an uncertainty-aware clustering method.
% but the quality of the result relies heavily on the initial random partitioning of global depth ranks and requires recomputation of data depth within each cluster.

\subsection{Probability Distribution Methods}
Modeling confidence intervals and probability density functions is another effective approach to address the challenge.
A number of methods study the spatial domain to estimate the probability distribution.
The probability plot of contours~\cite{kumpfVisualizingConfidenceClusterbased2018} enables per-point probability computation on each grid point of the spatial domain using signed distance fields.
EnconVis---a grid point-based technique---allows for probability band computation modeled with kernel density estimation~\cite{zhangEnConVisUnifiedFramework2023}. This method supports several other visual representations in its framework.
Probabilistic distribution function (PDF) and cumulative distribution function (CDF) bands are available for visualizing 2D contours~\cite{Pfaffelmoser2013}.
General collections of 2D function graphs are transformed into smooth probability maps using curve density estimations~\cite{lampeCurveDensityEstimates2011}. 
Kernel density estimation is generalized from points to polygonal surface patches for ensembles of surfaces, i.e., 3D contours~\cite{heEFESTAEnsembleFeature2019}. 

The probability distribution and confidence intervals can be analyzed using latent spaces.
The variability plot of streamlines~\cite{ferstlStreamlineVariabilityPlots2016} and contours~\cite{ferstlVisualAnalysisSpatial2016,ferstlTimehierarchicalClusteringVisualization2017} compute probabilistic bands of features by drawing ellipsoid regions in latent spaces created by the PCA and projecting them back to the spatial domain.
One benefit of these techniques is that they readily support the clustering of ensemble members in latent spaces by placing multiple ellipsoids.  

A previously overlooked limitation of these works is that their results do not align with those from data-depth methods.
One reason is that these approaches either rely on local spatial information or linear latent spaces, which fail to capture the complex nonlinear information of ensemble members that data depth methods effectively represent.
To address this, a preliminary exploration of visualizing ensemble members with VAE supports the estimation of probabilistic density of members and confidence bands~\cite{wu2025ensemblevisualizationvariationalautoencoder}.

% \zl{Delineate our work from Westermann's (cvp, cpp)--probability+PCA(kind of latent space) and contour boxplots--order stat}
Our new method is different from the aforementioned works in several aspects.
Our probabilistic modeling supports the analysis of multimodal ensembles, including tasks of data depth computation, member clustering, and probabilistic function estimation in the same latent space for consistency.
The use of a VAE provides improved density estimation that is more coherent with the ensemble member distributions represented using spaghetti plots or boxplots compared to existing methods.

\section{Ensemble Data Analysis with Latent Space Modeling}
\label{sec:methods}
Consider an ensemble of $N$ spatial features, such as contours, defined on the $m$-dimensional spatial domain ($m=2$ for images).
Each original member---for example, the green contour illustrated in~\cref{fig:workflow}---is converted into a high-dimensional representation in a feature space $X$ (\cref{sec:featurespace}).
We denote the resulting feature space representations as $\Xi=\{x_1,x_2,\cdots,x_N\}$, where each $x_i \in X$ is the object that our method operates on.
We assume that $X$ contains an unknown probability distribution $P(x)$ from which the observed feature representations $\{x_1, x_2,\cdots, x_N\}$ are drawn as samples.
The goal of ensemble visualization is to faithfully reveal the structure of $P(x)$ along with the members $\Xi$---analogous to how violin plots and boxplots show data distributions, with jittered data points for the observed samples.

However, since $X$ is of high dimensionality, its structure is hard to describe and analyze directly.
To address this, we use the encoder of a VAE to map each $x_i \in X$ into a low-dimensional latent space $Z \subset \mathbb{R}^k$ with $k \ll \dim(X)$.
Rather than collapsing $x_i$ to a single deterministic point, the encoder maps it to a probability distribution $z_i$ in $Z$, thereby preserving the uncertainty inherent in the encoding.
This process is governed by the conditional probability relationship:
\begin{equation}
    p(x)p(z \mid x) = p(x \mid z)p(z)\;,
    \label{eqn:basemodel}
\end{equation}
where $p(z)$ is the prior density in the latent space.
The encoder approximates the posterior $p(z \mid x)$ as a Gaussian distribution, encoding each feature into a distributional latent representation.
The decoder models $p(x \mid z)$ to reconstruct features from latent samples.
Because the VAE is trained to minimize reconstruction error, the latent space $Z$ retains the principal structure of $X$ while reducing the dimensionality to a level amenable to efficient statistical analysis.

As shown in the workflow (\cref{fig:workflow}), our method supports data depth computation, uncertainty-aware clustering, and the estimation of probabilistic density---all derived from the same latent space for consistency.

\subsection{Probabilistic Modeling with VAE}
\label{sec:vae_method}
We detail the construction of feature space and latent space with a VAE, and the probabilistic similarity measurement that serves as the basis for subsequent analysis tasks.
\subsubsection{Feature Space Creation}
\label{sec:featurespace}
\begin{figure}[t]
  \centering
  \includegraphics[width=0.8\linewidth]{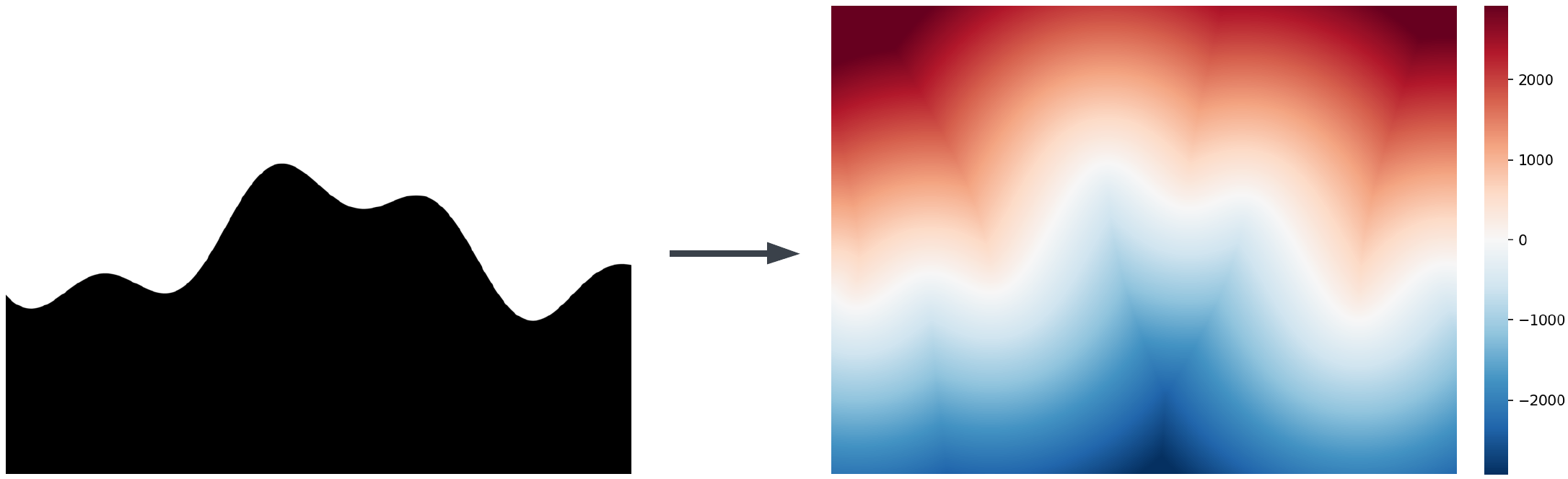}
  \caption{
    The signed distance field (right) of a binary contour mask of an ensemble member (left). Positive values (exterior of the region) are in red, the zero level set (the original contour) is in white, and negative values (interior) are in blue.}
  \label{fig:sdt}
\end{figure}

We first apply the signed distance transform to convert each ensemble member into a signed distance field (SDF) image. 
Each contour is represented as the boundary of a binary feature region (\cref{fig:sdt}(left)), which is obtained either by thresholding an input scalar field, i.e., 1 for regions with scalar values greater than the isovalue, and 0 otherwise, or directly from an input segmentation mask (1 for interior, 0 for exterior).
Let $\Omega_i$ denote the feature region of the $i$-th member and $\partial \Omega_i$ its boundary.
For each spatial location $u$, its SDF image value is defined as:

\begin{equation}
  x_i(u)=
  \begin{cases}
    -d\!\left(u,\partial \Omega_i\right), & u\in \Omega_i,\\
    \ \ d\!\left(u,\partial \Omega_i\right), & u\notin \Omega_i,
  \end{cases}
\end{equation}
where $d(\cdot,\partial \Omega_i)$ is the Euclidean distance to the contour. 
Therefore, the SDF value is negative inside the feature region, positive outside, and zero on the contour boundary, as shown in~\cref{fig:sdt}. 
Compared to a binary encoding, this representation preserves boundary geometry in a continuous manner and provides richer local variation near the contour, which is important for learning shape-aware latent features.
\rev{In our implementation, the SDF is computed on the native grid of the input scalar field or segmentation mask, i.e., each grid point in the original data is assigned a signed distance value to the contour.}

The feature space $X$ is constructed by discretizing each member on a regular grid of size $(w,h)$ and reshaping it into a vector in $\mathbb{R}^{w\times h}$. 
Subsequently, all samples are normalized by a global $z$-score computed over all pixels of all ensemble members so that the sign structure of the SDF is preserved while the numerical scale is standardized as the input of the neural network.

\subsubsection{Latent Space Construction with the VAE}
We model the feature space with a convolutional VAE~\cite{kingmaAutoencodingVariationalBayes2022}.
The encoder progressively compresses the input SDF image into a 
compact feature tensor through a series of strided convolutional 
blocks, and is then mapped by two fully connected heads to the 
mean vector $\mu_i$ and log-variance vector $\log \sigma_i^2$ of
a diagonal Gaussian that approximates the posterior:
\begin{equation}
  p(z\mid x_i)\approx\mathcal{N}\!\left(\mu_i,\mathrm{diag}(\sigma_i^2)\right)\;.
\end{equation}
\rev{The same network configuration is used for all datasets in this paper; its implementation and model-adequacy analysis are described in~\cref{sec:implement}.}

The latent variable $z_i$ is sampled with the reparameterization trick, and the decoder first projects $z_i$ back to the compressed tensor space and then reconstructs the input resolution by alternating bilinear upsampling and convolution. This design preserves spatial coherence during reconstruction while learning a low-dimensional probabilistic embedding of each ensemble member. 

The VAE is trained by minimizing the VAE objective: 
\begin{equation}
  \mathcal{L}
  =
  \frac{1}{N}\sum_{i=1}^{N}
  \left(
    \left\|x_i-\hat{x}_i\right\|_2^2
    +
    \beta\,
    D_{\mathrm{KL}}\!\left(
      p(z\mid x_i)\,\|\,\mathcal{N}(0,I)
    \right)
  \right),
\end{equation}
where the first term is the reconstruction loss and the second term regularizes the latent distribution toward the standard normal prior. In practice, we use mean squared error for reconstruction and optimize the network with Adam~\cite{kingma2017adam}. 
To avoid unstable optimization in the early stage, the KL term is linearly annealed from $0$ to the target weight $\beta$ during a warm-up period, and the learning rate is further controlled by a cosine annealing schedule. 

\subsubsection{Similarity Measurement}
After training, each ensemble member $x_i$ is encoded as a Gaussian distribution $\mathcal{N}(\mu_i, \mathrm{diag}(\sigma_i^2))$ in $Z$, where the posterior mean $\mu_i$ captures the member's structural identity and the posterior variance $\sigma_i^2$ reflects the associated encoding uncertainty.
We retain this full distributional representation to define a probabilistic similarity between pairs of members.

For two members $x_i$ and $x_j$ with latent representations $z_i \sim p(z\mid x_i)$ and $z_j \sim p(z\mid x_j)$, we measure their similarity by the probability density that these two latent variables coincide~\cite{shiProbabilisticFaceEmbeddings2019}:
\begin{equation}
    p(z_i = z_j) = \int p(z_i \mid x_i)\, p(z_j \mid x_j)\, \delta(z_i - z_j)\, dz_i\, dz_j\;,
    \label{eqn:pzij}
\end{equation}
where $\delta(\cdot)$ is the Dirac delta function.
\begin{figure}[htb]
    \centering
    \includegraphics[width=0.9\linewidth]{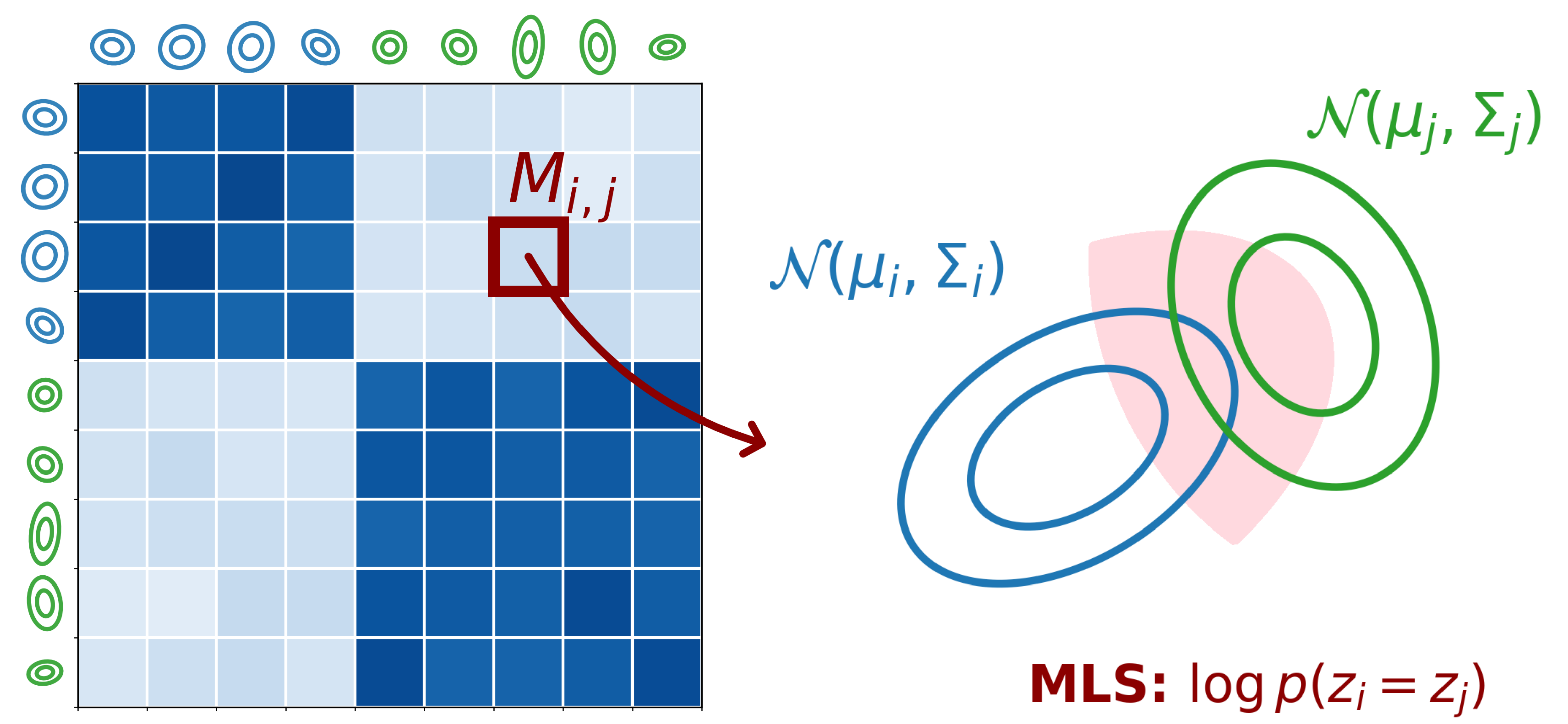}
    \caption{The pairwise MLS similarity matrix $\mathbf{M}$ is an $N\times N$ symmetric matrix. Left: Each cell $\mathbf{M}_{i,j}$ encodes the mutual likelihood score between ensemble members $i$ and $j$. Marginal distributions along the axes depict the latent Gaussian distribution of each member, colored by cluster assignment. Right: a conceptual view of the MLS computation in the latent space, where the similarity is determined by the overlap (shaded region) between two Gaussian distributions $\mathcal{N}(\mu_i, \Sigma_i)$ and $\mathcal{N}(\mu_j, \Sigma_j)$.}
    \label{fig:mlsmatrix}
\end{figure}
The log-likelihood solution of~\cref{eqn:pzij} is given by the mutual likelihood score (MLS):
\begin{equation}
\text{MLS} = -\frac{1}{2}\sum_{l=1}^{k} \left( \frac{(\mu_i(l)-\mu_j(l))^2}{\sigma_i^2(l)+\sigma_j^2(l)} + \log\left(\sigma_i^2(l)+\sigma_j^2(l)\right) \right) - C\;,
\label{eqn:MLS}
\end{equation}
where \(C = \frac{k}{2}\log 2\pi\), \(\mu_i(l)\) denotes the \(l\)-th dimension of \(\mu_i\), and \(\sigma_i(l)\) denotes the \(l\)-th dimension of \(\sigma_i\).
Details of the derivation are documented in Sec. 2 of the supplemental material~\cite{vaeEnVisFrameworkSuppl}.

To efficiently support the subsequent analytical tasks, we pre-compute the pairwise MLS for all members in the ensemble.
This results in an $N \times N$ symmetric similarity matrix $\mathbf{M}$, where each element $\mathbf{M}_{i,j}$ is defined as:
\begin{equation}
\mathbf{M}_{i,j} = \text{MLS}(z_i, z_j)\;.
\end{equation}

As shown in~\cref{fig:mlsmatrix}, each ensemble member is encoded as a Gaussian distribution in the latent space (elliptical contours along the matrix margins). The entry $\mathbf{M}_{i,j}$ quantifies the similarity between members $i$ and $j$ by measuring the overlap of their respective distributions $\mathcal{N}(\mu_i, \Sigma_i)$ and $\mathcal{N}(\mu_j, \Sigma_j)$ in the latent space (right). Members belonging to the same cluster exhibit high MLS values (dark cells), while cross-cluster pairs yield low values (light cells), revealing the underlying group structure. This pre-computed MLS matrix serves as the unified foundation for both the uncertainty-aware clustering (\cref{sec:cluster}) and the latent data depth computation (\cref{sec:datadepth}), ensuring that the uncertainty is consistently preserved and computationally efficient across our framework.

\subsection{Uncertainty-Aware Clustering}
\label{sec:cluster}

For ensembles of multimodal distributions, partitioning the data into distinct clusters allows each subset to be approximated as a unimodal distribution. 
This preprocessing step extends the applicability of tools originally designed for unimodal data---such as latent space data depth (\cref{sec:datadepth}) and probabilistic density plots (\cref{sec:density})---to arbitrary ensemble distributions.

\begin{figure}[htb]
    \centering
    \begin{subfigure}[b]{0.32\linewidth}
        \centering
        \includegraphics[width=\linewidth]{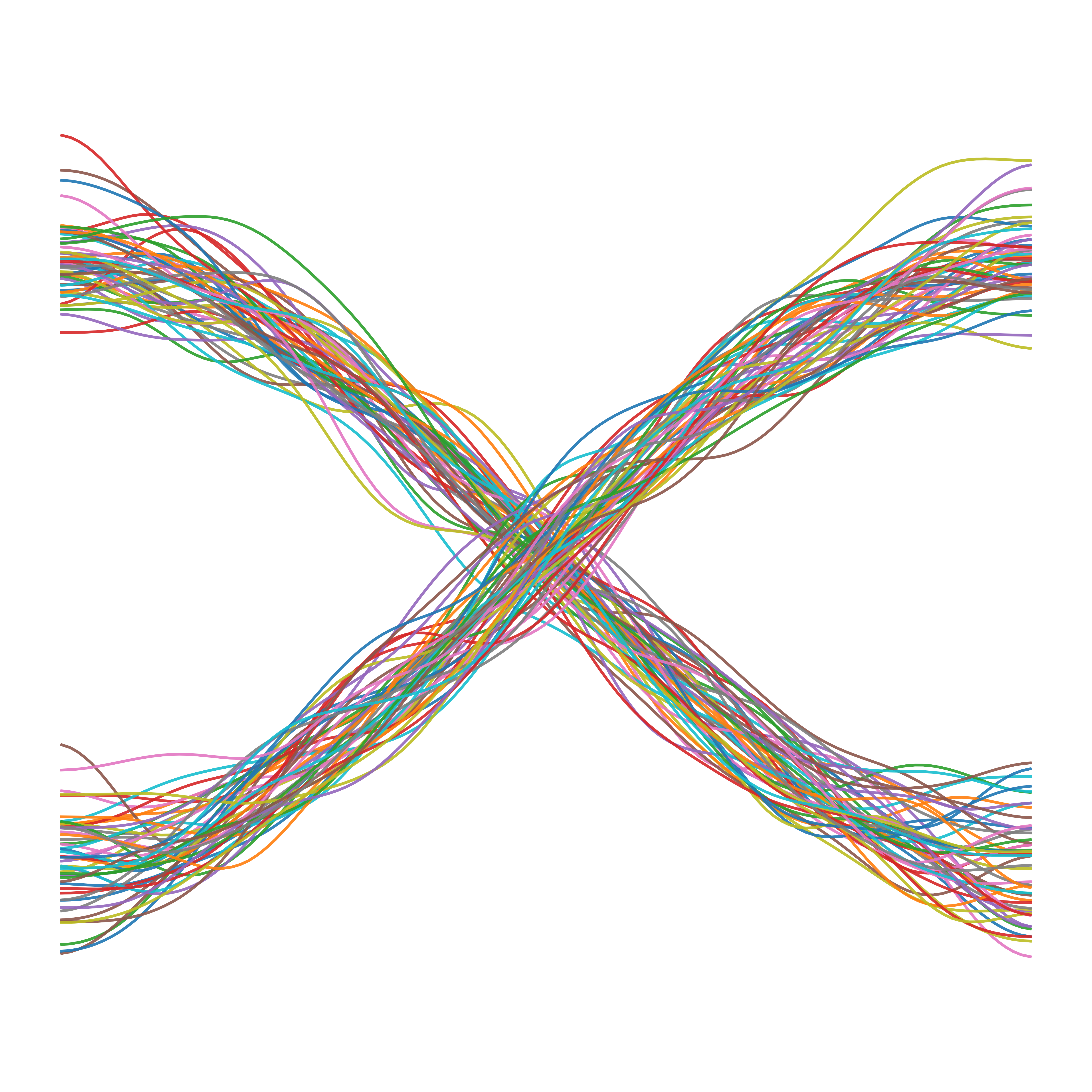}
        \caption{Synthetic data}
        \label{fig:clusterProblem_a}
    \end{subfigure}\hfill
    \begin{subfigure}[b]{0.32\linewidth}
        \centering
        \includegraphics[width=\linewidth]{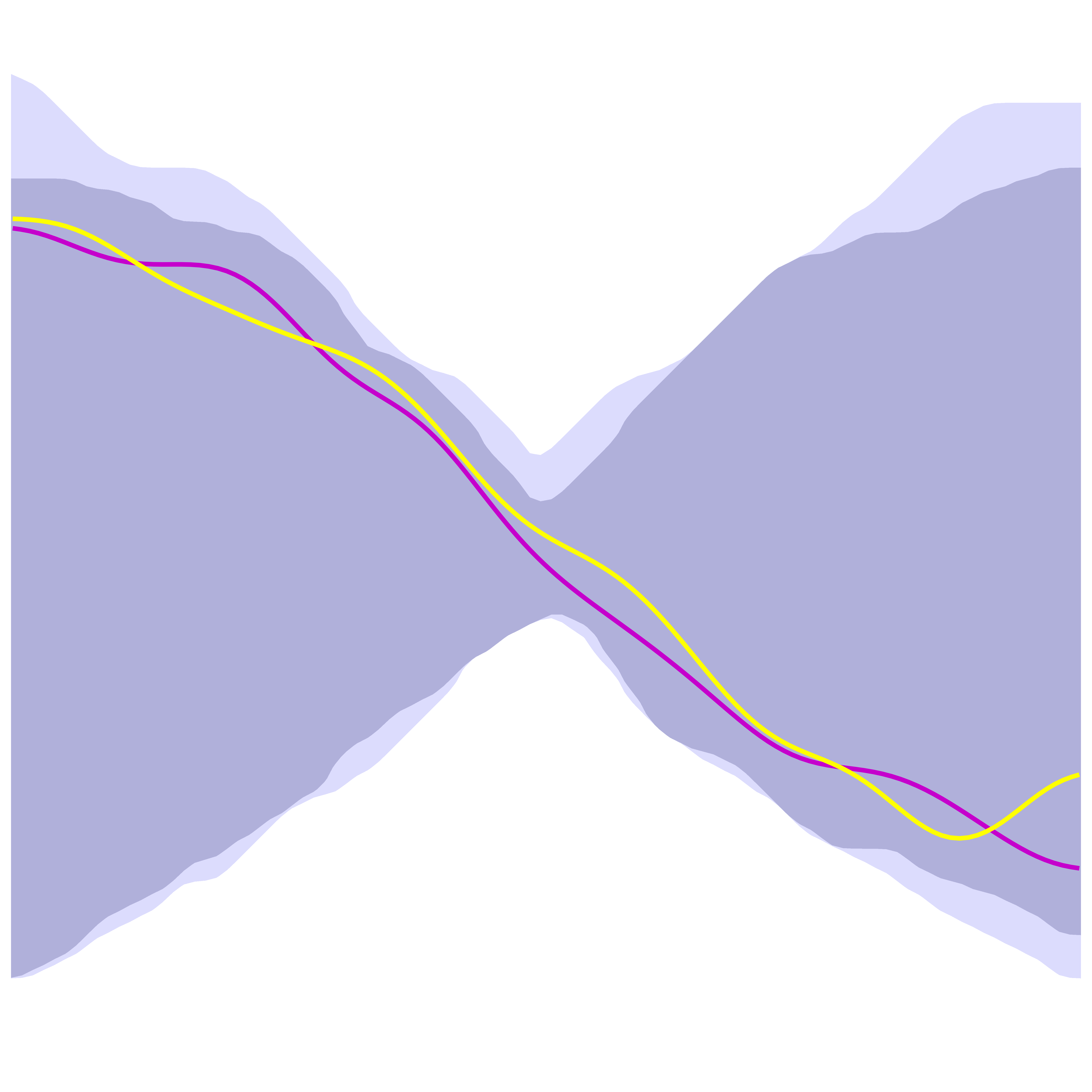}
        \caption{The boxplot of all samples}
        \label{fig:clusterProblem_b}
    \end{subfigure}\hfill
    \begin{subfigure}[b]{0.32\linewidth}
        \centering
        \includegraphics[width=\linewidth]{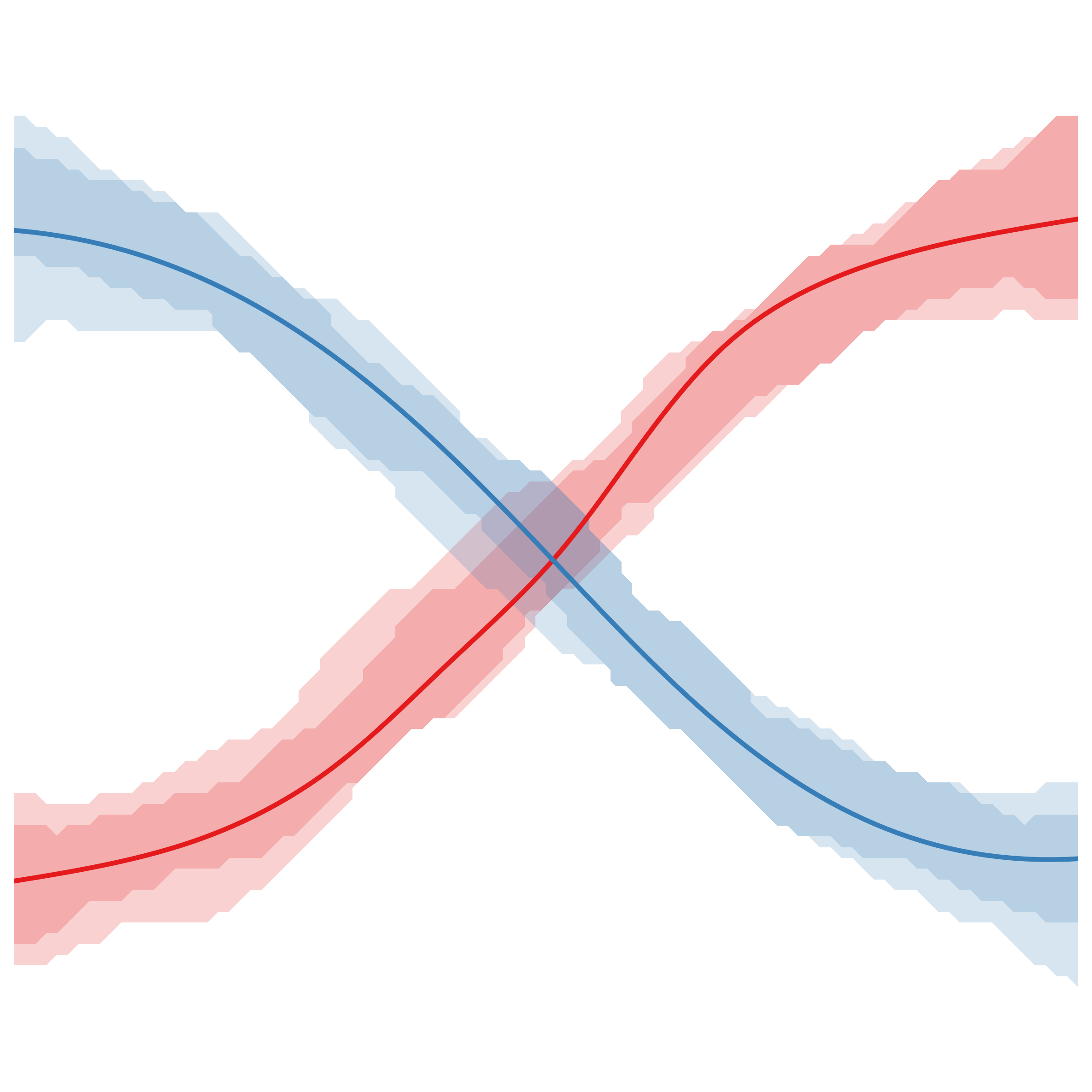}
        \caption{Boxplots of clustered samples}
        \label{fig:clusterProblem_c}
    \end{subfigure}
    \caption{A synthetic multimodal ensemble data (a) is visualized using boxplots for (b) all members under the unimodal assumption, and (c) two groups of members after our uncertainty-aware clustering.}
    \label{fig:clusterProblem}
\end{figure}
An erroneous example of applying data depth under the unimodal assumption to a synthetic multimodal ensemble is illustrated in~\cref{fig:clusterProblem}(b): regions that are not covered by ensemble samples (see the spaghetti plot in~\cref{fig:clusterProblem}(a)) are falsely identified as valid regions within the $50\%$ population band (dark purple). 
The generation of this cross-shaped ensemble is detailed in Sec. 7.2 of the supplemental material. 

Agglomerative hierarchical clustering (AHC) is a widely used bottom-up clustering technique~\cite{johnson1967hierarchical}. 
It initializes each sample as an individual cluster and iteratively merges the most similar pair of clusters until a specified number of clusters is reached. Specifically, the Ward linkage method in traditional AHC merges the cluster pair $(A^*, B^*)$ that minimally increases the overall error sum of squares (ESS):
\begin{align}
(A^*,B^*) &= \arg\min_{A,B} \Delta \text{ESS}(A,B)\;, \label{eqn:mergeAHC}    \\
\text{with\space} \text{ESS} &= \sum_{i=1}^{K} \sum_{z \in C_i} \|z - \bar{z}_i\|^2\;,
\label{eqn:ESS}
\end{align}
where $A$ and $B$ are any two active clusters, and $\bar{z}_i$ is the centroid of cluster $C_i$.

\begin{figure}[htb]
    \centering
    \includegraphics[width=0.9\linewidth]{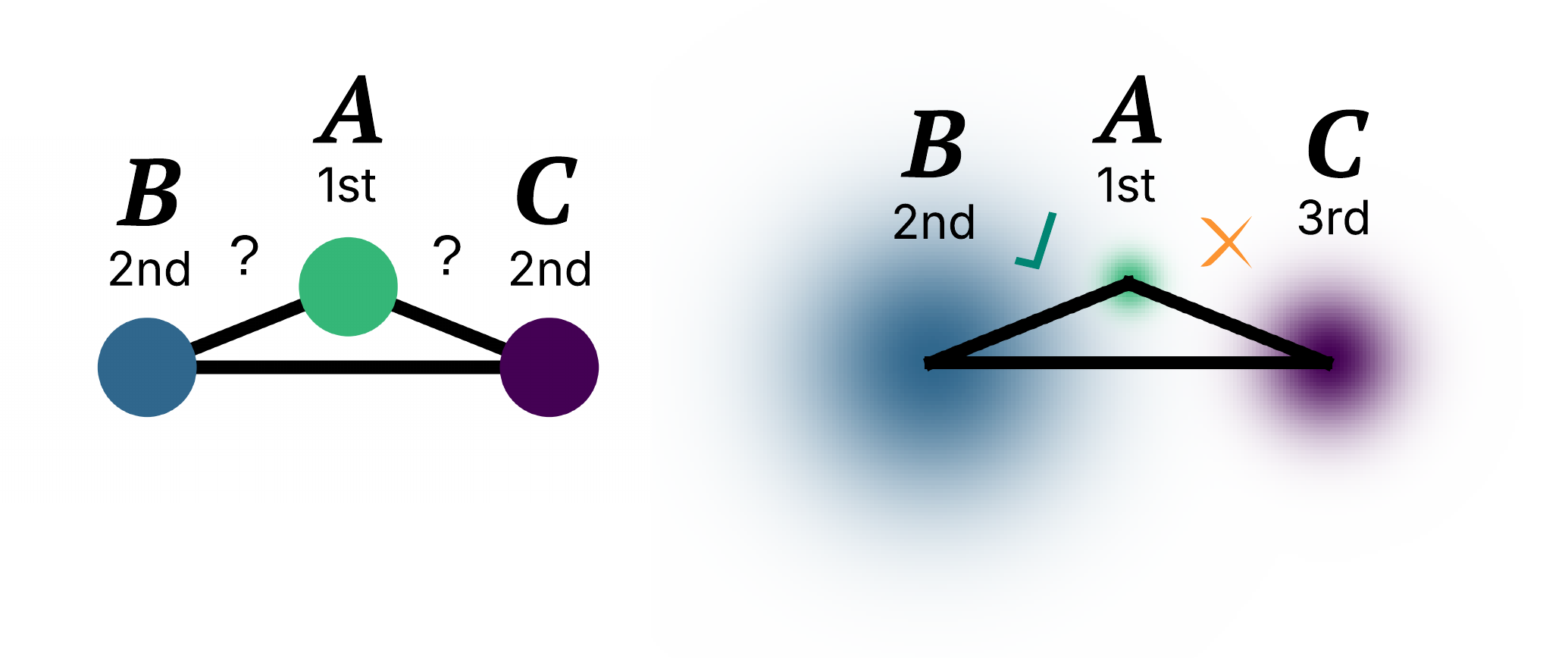}
    \vspace{-2em}
    \caption{
An ambiguity exists for the traditional AHC for three equal-
distanced deterministic data points, as point A can be clustered with either B or C. The Mahalanobis depth assigns points B and C the same rank (left). For three distributions with distinct uncertainties, although the centers of B and C remain equidistant from A,  A is clustered with point B as it has a higher probability compared to point C, and our MLS depth yields different depth values because the distributional overlaps differ (right).} 
    \label{fig:depthtoy}
\end{figure}

However, traditional AHC relies solely on deterministic distance metrics (e.g., Euclidean distance), which ignores the probabilistic nature of our latent space. In our framework, each ensemble member $x$ is mapped to a conditional distribution $P(z\mid x)$, inherently containing uncertainty information characterized by its variance. The importance of incorporating this uncertainty is illustrated in~\cref{fig:depthtoy}. Given three equidistant points, distance-based AHC randomly assigns point $A$ to either $B$ or $C$ (\cref{fig:depthtoy}(left)). Conversely, when distributional uncertainty is considered, point $A$ correctly clusters with point $B$ because it exhibits a higher probability overlap (\cref{fig:depthtoy}(right)).

To integrate uncertainty into the clustering process, we replace the distance-based ESS with a similarity-based metric derived from the Mutual Likelihood Score (MLS). Using the pre-computed symmetric MLS matrix $\mathbf{M}$ introduced in \cref{sec:vae_method}, we define the Summed MLS (SMLS) as the negative sum of all pairwise similarities within each cluster:
\begin{equation}
\text{SMLS} = -\sum_{i=1}^{K} \sum_{p,q \in C_i} \mathbf{M}_{p,q}\;.
\label{eqn:SMLS}
\end{equation}
By replacing $\Delta \text{ESS}$ with $\Delta \text{SMLS}$ as the merging criterion, the target pair to merge becomes:
\begin{equation}
(A^*,B^*) = \arg\min_{A,B} \Delta \text{SMLS}(A,B)\;.
\end{equation}

This formulation leads to our uncertainty-aware clustering algorithm, namely, MLS-AHC. The detailed procedure is outlined in~\cref{alg:uncertainAHC}. 
Notably, since the MLS matrix $\mathbf{M}$ is pre-computed, step 2 of the algorithm requires no additional mathematical operations, making MLS-AHC efficient.
\begin{algorithm}[ht]
\caption{Mutual Likelihood Score AHC (MLS-AHC)}
\label{alg:uncertainAHC}
\begin{algorithmic}[1]
\Require Pre-computed symmetric MLS matrix $\mathbf{M} \in \mathbb{R}^{N \times N}$ (\cref{sec:vae_method}), desired clusters $K$

\vspace{6pt}
\State // Initialize: each sample starts as its own cluster
\For{$i=1$ \textbf{to} $N$}
    \State $C_i \gets \{i\}$, \quad $label[i] \gets i$
\EndFor

\vspace{6pt}
\State // Clustering loop
\While{number of active clusters $> K$}
    \For{each pair of active clusters $(C_i, C_j)$}
        \State Compute: $\text{SMLS}(C_i,C_j) \gets -\!\sum_{p\in C_i}\sum_{q\in C_j} \mathbf{M}_{p,q}$
    \EndFor

    \vspace{4pt}
    \State $(i^*,j^*) \gets \arg\min_{i\neq j} \,\text{SMLS}(C_i,C_j)$

    \vspace{4pt}
    \State // Merge clusters with the minimal SMLS increment
    \State $C_{i^*} \gets C_{i^*} \cup C_{j^*}$
    \State $\forall\, p\in C_{j^*},\; label[p] \gets i^*$
    \State Remove cluster $C_{j^*}$
\EndWhile

\vspace{6pt}
\State Relabel clusters sequentially to $\{0,1,\dots,K\!-\!1\}$
\State \Return $label$
\end{algorithmic}
\end{algorithm}

The result of MLS-AHC on the synthetic data (\cref{fig:clusterProblem}(a)) are shown in~\cref{fig:clusterProblem}(c).
Compared to the boxplot of all samples (\cref{fig:clusterProblem}(b)), boxplots of clustered samples correctly visualize the two modes (blue and red) of the ensemble distribution (\cref{fig:clusterProblem}(c)), where the median members are highlighted using solid lines.
\newcommand{\tumorheight}{2.5cm}
\begin{figure}[htb]
\centering
\subfloat[Spaghetti plot]{\includegraphics[height=\tumorheight]{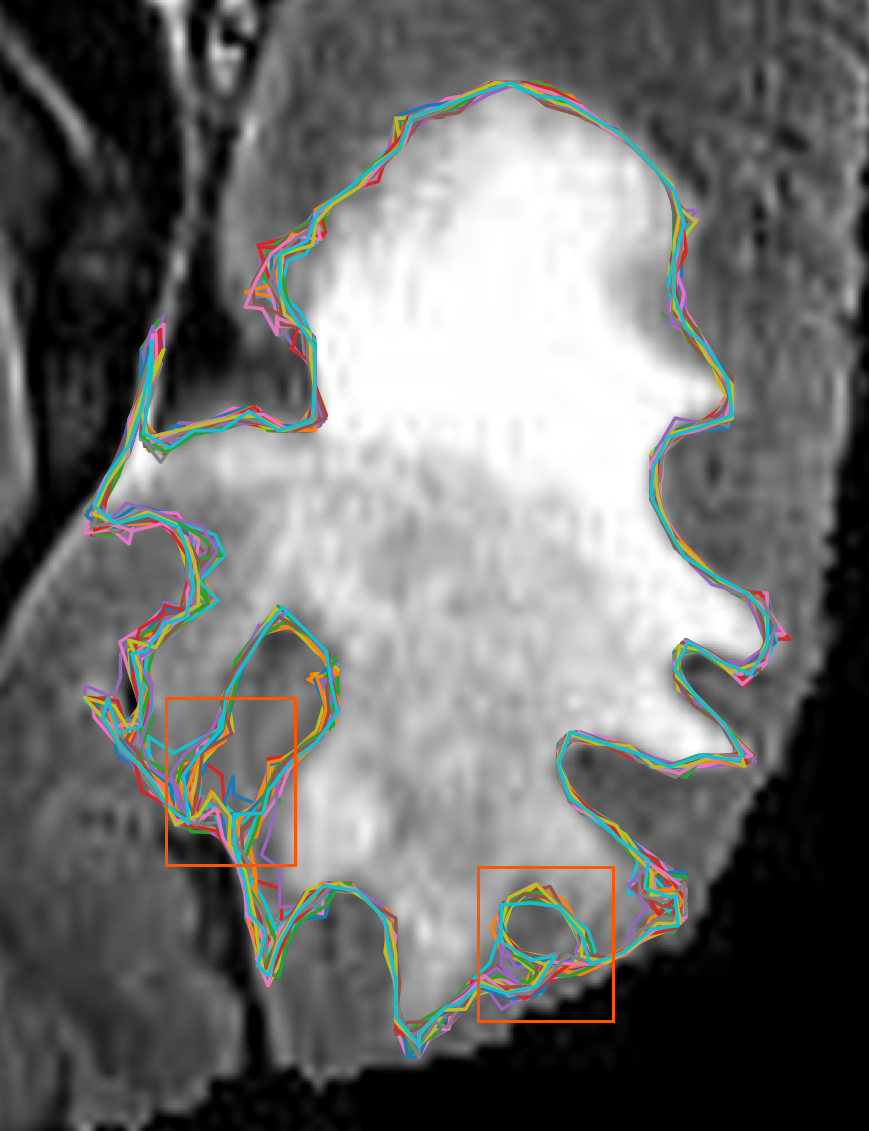}}
\subfloat[Cluster 1]{\includegraphics[height=\tumorheight]{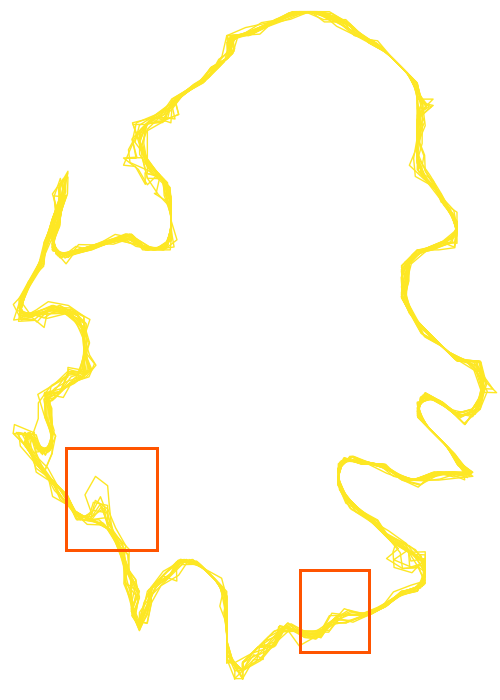}}
\subfloat[Cluster 2]{\includegraphics[height=\tumorheight]{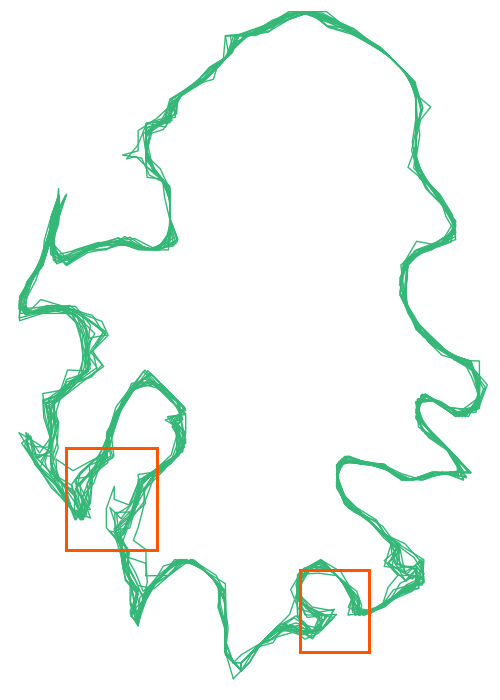}}
\subfloat[Cluster 3]{\includegraphics[height=\tumorheight]{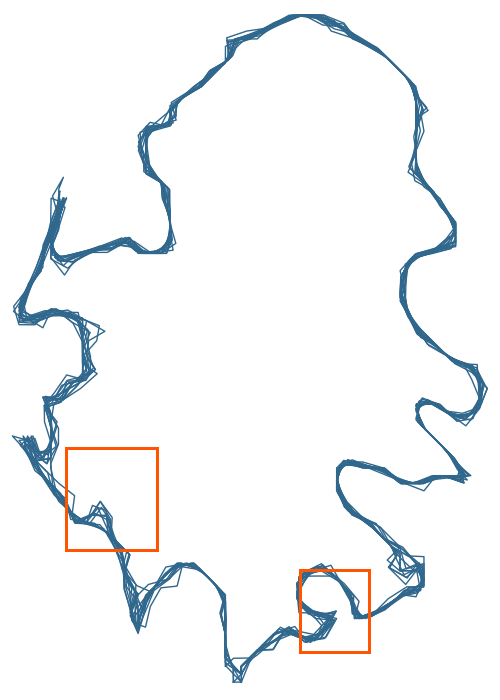}}
\subfloat[Cluster 4]{\includegraphics[height=\tumorheight]{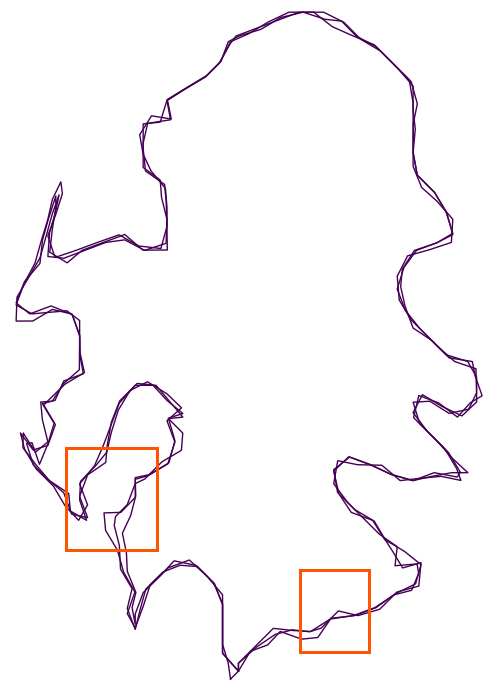}}
 \caption{A tumor segmentation contour ensemble is clustered with MLS-AHC (a). Clusters (b--e) differ in the orange-boxed regions of the tumor. }
 \vspace{-1em}
 \label{fig:tumorClusters}
\end{figure}

Results of MLS-AHC are further demonstrated with two real-world ensembles.
An ensemble of brain tumor segmentation results, represented by their extracted segmentation contours (\cref{fig:tumorClusters}(a)), is partitioned into four clusters.
The resulting groups (\cref{fig:tumorClusters}(b--e)) exhibit distinct lesion patterns, differing primarily in the opening and closing of the bottom-left and bottom-right regions of the tumor boundary.
Note that for this ensemble, binary contour images are used as differences of the opennings and closings sometimes become invisible in SDF images.

A weather forecast ensemble is partitioned into two clusters ($K=2$), as shown in \cref{fig:teaser}.
The corresponding probability density plots (\cref{sec:density}) in \cref{fig:teaser} reveal two distinct flow patterns that are clearly separated by our method.
In both cases, MLS-AHC operates directly on the precomputed MLS matrix without any additional distance computation, making the clustering step itself negligible in cost.

\subsection{Data Depth in the Latent Space}
\label{sec:datadepth}

Following the center-outward view of data depth introduced earlier, we define data depth in the latent space so that contour boxplots can be constructed from the same probabilistic representation used for clustering and density estimation.
Traditionally, the Mahalanobis depth~\cite{mahalanobis1936generalized} is used to rank multidimensional data by computing the Mahalanobis distance from each member to the distribution mean $\mu$:
\begin{equation}
    D_M(z_i) = \frac{1}{1 + (z_i-\mu)^\top \Sigma^{-1} (z_i-\mu)}\;.
\end{equation}
In the isotropic latent space of a VAE, calculating the Mahalanobis depth simplifies to computing the Euclidean depth. 
Geometrically, Euclidean depth measures centrality based on average proximity. 
The Euclidean depth $D_E(z_i)$ can be formulated as inversely proportional to the average pairwise squared Euclidean distance between $z_i$ and all other members:
\begin{equation}
    D_E(z_i) \propto \left( \frac{1}{N} \sum_{j=1}^{N} \|z_i - z_j\|_2^2 \right)^{-1} \;.
\end{equation}
The connection between $D_M$ and $D_E$ is derived in Sec. 3.1 of the supplemental material.

From a probabilistic perspective, this geometric metric has a statistical interpretation. Specifically, the probability density of an isotropic Gaussian distribution decays exponentially with the squared distance from its mean. 
Therefore, the log-likelihood of observing any member $z_j$ from a center $z_i$ is proportional to the negative squared distance $-\|z_i - z_j\|_2^2$. 
Under the assumption of an isotropic Gaussian distribution, the negative squared Euclidean distance is strictly proportional to the log-likelihood. 
Therefore, minimizing the average squared distance is mathematically equivalent to maximizing the joint log-likelihood of the entire ensemble, assuming $z_i$ acts as the spatial center.

However, representing an ensemble member merely as a deterministic mean point discards the valuable uncertainty information encoded in its posterior variance.
In our probabilistic modeling, each member is not a single point, but a multivariate Gaussian distribution. Therefore, evaluating how ``central'' or ``typical'' a member is should depend not on deterministic point-to-point distances, but on the probabilistic overlap between distributions.

This distinction is illustrated in~\cref{fig:depthtoy}. When ensemble members are treated as deterministic points (\cref{fig:depthtoy}(left)), the Mahalanobis depth ranks $B$ and $C$ identically because they are equidistant from $A$. Once the encoding uncertainty is taken into account (\cref{fig:depthtoy}(right)), the distributions of $B$ and $C$ exhibit different degrees of overlap with $A$, and our MLS-based depth correctly distinguishes them. This motivates the probabilistic generalization of data depth described below.

We adapt the log-likelihood perspective to these probabilistic representations. 
Instead of computing the likelihood of how close deterministic points are, we calculate the expected log-likelihood that two latent distributions correspond to the same underlying feature. 
As introduced in~\cref{sec:vae_method}, this distribution-level log-likelihood is explicitly captured by the MLS (\cref{eqn:MLS}).

Consequently, the joint log-likelihood of member $i$ acting as the probabilistic center of the entire ensemble is proportional to the average of its pairwise MLS values. 
This allows us to define the probabilistic latent data depth $D(z_i)$ directly through MLS:
\begin{equation}
    D(z_i) = \frac{1}{N} \sum_{j=1}^{N} \text{MLS}(z_i, z_j)\;.
\end{equation}
By replacing the deterministic distance metric with the probabilistic log-likelihood, this formulation seamlessly accounts for both the mean and the variance. 
The deepest member is exactly the one that maximizes the mutual likelihood across the ensemble, inherently preserving generative uncertainty within an objective statistical framework.
The isotropic Gaussian case, where the Euclidean depth is equivalent to maximum-likelihood ranking, is derived in Sec.~3.2 of the supplemental material.

Importantly, because $\mathbf{M}$ is readily pre-computed (\cref{sec:vae_method}), we can directly reuse it to obtain the data depth. For any member $i$, its depth $D(z_i)$ is simply the row average of matrix $\mathbf{M}$:
\begin{equation}
    D(z_i) = \frac{1}{N} \sum_{j=1}^{N} \mathbf{M}_{i,j}\;.
\end{equation}
As a result, this generalized depth introduces no additional computational overhead to our method.

\begin{figure*}[tb]
    \centering
        \subfloat[Hippocampus segmentation ensemble]{\includegraphics[height=2.4cm]{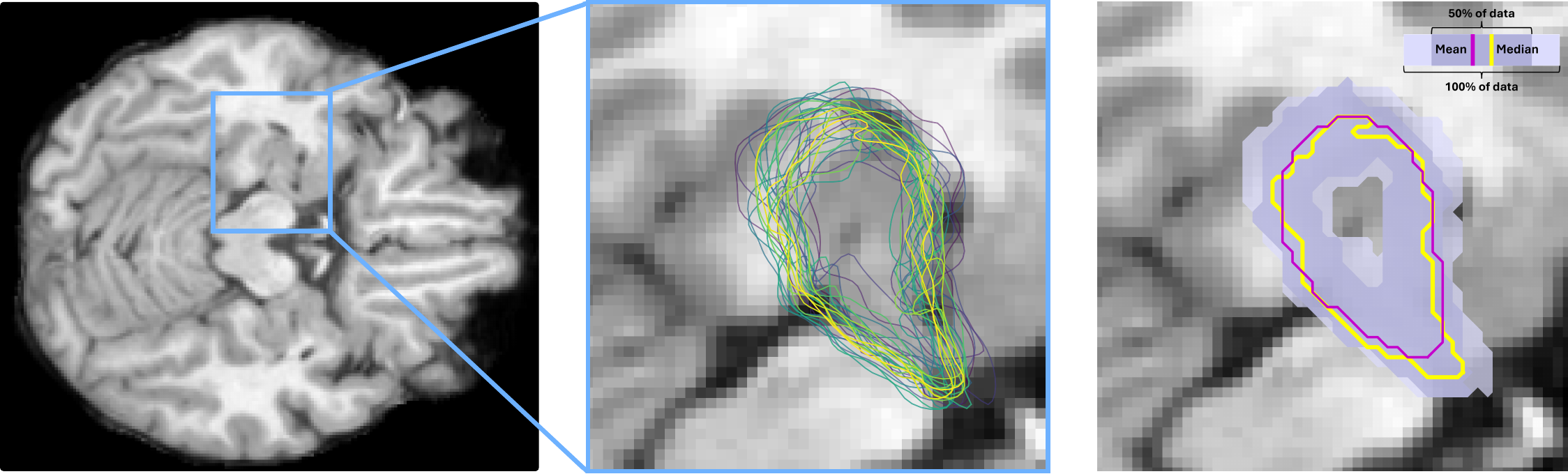}}    \hspace{1em}
    \subfloat[Third and fourth ventricles segmentation ensemble]{\includegraphics[height=2.4cm]{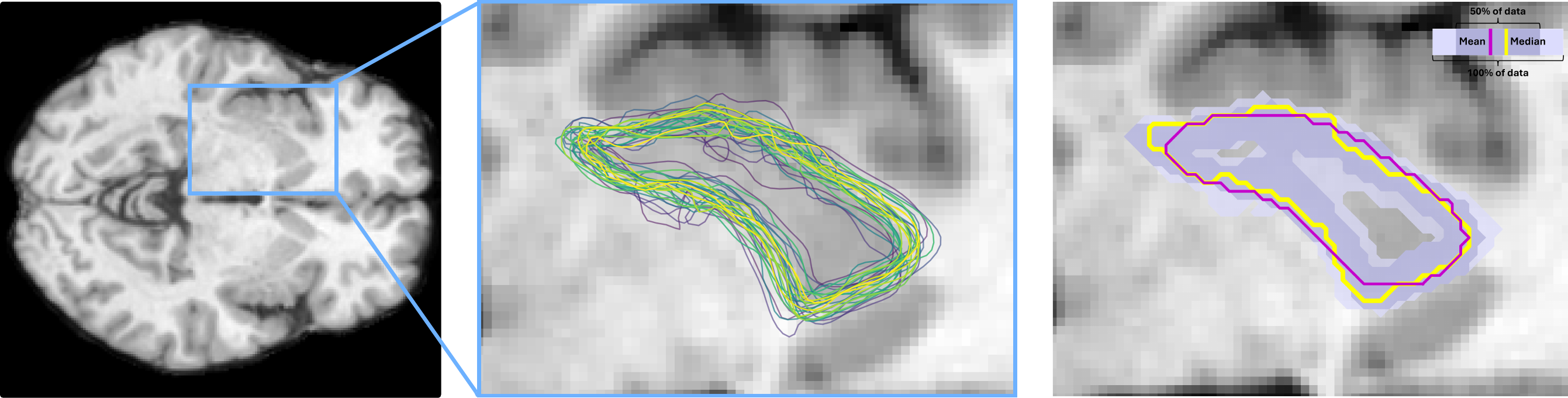}}   
    \vspace{-1ex}
    \caption{
    Latent space data depth applied to two segmentation ensembles---each of 30 members---from the IXI dataset~\cite{antonelli2022medical}.
    Results of the hippocampus and third and fourth ventricles are shown in (a) and (b), respectively.
    For each result, left: the original brain MRI slice, with the ventricular region highlighted in a blue box. Middle: a spaghetti plot of all segmentation contours, colored by normalized depth rank using the viridis colormap (yellow for the deepest, purple for the shallowest). Right: the contour boxplot derived from the depth ranking.
    }\label{fig:haimaDepth}
\vspace{-1ex}
\end{figure*}

An example of the latent space data depth applied to a real-world medical ensemble is shown in~\cref{fig:haimaDepth}(a).
The ensemble comprises 30 hippocampus segmentation contours from the IXI dataset (see~\cref{sec:ixidata} for details).
In the spaghetti plot (\cref{fig:haimaDepth}(a), middle), each member is colored according to its normalized depth rank: members closer to the distributional center appear in yellow, while those at the periphery appear in purple.
The depth ranking induces a contour boxplot (\cref{fig:haimaDepth}(c), right), which summarizes the ensemble variability in a compact visual form.
Specifically, the light purple band covers the region enclosed by all members (100\% of the depth-ranked population), while the dark purple band narrows to the top 50\%, highlighting the area of highest consensus.
The yellow line marks the median contour---the member with the greatest depth, representing the most centrally located segmentation---and the magenta line is the mean contour decoded from the latent space centroid $\bar{\mu} = \frac{1}{N}\sum_{i=1}^{N}\mu_i$ as detailed in Sec. 6 of the supplemental material.
The visualization provides an effective overview of the variability of hippocampal boundary delineations.

Furthermore, the same pre-computed matrix $\mathbf{M}$ underpins both data depth and the uncertainty-aware clustering described in \cref{sec:cluster}. When users adjust the desired number of clusters, the clustering algorithm operates entirely on $\mathbf{M}$ without recomputing any pairwise likelihoods; once a new partition is obtained, the within-cluster depth of each member reduces to averaging the corresponding submatrix rows. Both steps involve only index look-ups and summations over the existing matrix entries, incurring negligible cost regardless of the number of clusters chosen. This efficiency enables rapid, interactive exploration of different clustering configurations and their associated depth rankings.

\section{Coherent Ensemble Visualization}
\label{sec:uncertainLobes}
An important benefit of using a VAE for ensemble visualization is its power to approximate complex functions, such as the contour members.
Therefore, we devise a VAE-based probability density plot visualization method.
In this section, we elaborate on this visualization technique and cover implementation details of our overall method.

\subsection{Probability Density Plots}
\label{sec:density}
Beyond discrete order statistics and clustering, a continuous representation of the probability distribution of contours in the spatial domain is also important.
To this end, we leverage the learned posterior distributions in the latent space to generate a large number of synthetic contours, from which a smooth density field is constructed.
These contours serve as Monte Carlo samples for estimating a continuous spatial density field beyond the sparse trajectories of the observed members.

The core idea is dense posterior sampling: a member $i$ is randomly drawn from the ensemble, and a latent sample $z$ is generated from its posterior $z \sim \mathcal{N}(\mu_i, \mathrm{diag}(\sigma^2_i))$, where $\mu_i$ and $\sigma^2_i$ are the mean and variance output by the encoder (\cref{sec:vae_method}).
This process is repeated to obtain a dense set of latent samples that provides an empirical approximation to the underlying distribution of the observed ensemble.
Each latent sample is then decoded by the VAE back into the spatial domain as an SDF image.
Because the SDF is a continuous scalar field (\cref{sec:featurespace}), its zero-level set is inherently a well-defined contour; we extract it efficiently using the marching squares algorithm.
This continuity guarantee ensures that every decoded sample produces a geometrically valid contour, preserving the structural integrity of the generated members.
Individual synthetic contours are decoded approximations rather than physical realizations, and their fidelity depends on the VAE reconstruction and latent representation; we therefore validate their collective use through the induced density fields, based on agreement with observed contours, recovery of the two known X-shaped branches, and depth--density coherence (\cref{sec:evalDensity}).
For clusters partitioned by MLS-AHC, the same sampling procedure is applied independently within each cluster, yielding a per-cluster density field that captures the spatial distribution specific to that sub-population.

To construct the density map for a given cluster, the extracted contours are rasterized line segment by line segment onto a high-resolution target grid to accumulate spatial hit counts.
A spatial smoothing step is then applied to transform the sparse, discrete hit counts into a smooth and continuous density field.
Specifically, a compact 2D smoothing kernel of prescribed radius is constructed and convolved with the hit-count grid using the fast Fourier transform (FFT), which reduces the computational complexity from $O(G\!\cdot\!K)$ for a na{\"i}ve per-pixel loop (where $G$ is the number of grid cells and $K$ the kernel area) to $O(G\log G)$, independent of the kernel size.
In practice, the FFT approach accelerates density computation by one to two orders of magnitude compared to the direct spatial convolution.

The resulting density plots (\cref{fig:teaser}(c),~\cref{fig:densityEval}(a), and~\cref{fig:scalarflow}) effectively visualize the spatial uncertainty within each cluster, where high-density regions indicate a greater likelihood of contour occurrence.
The two clusters identified by MLS-AHC yield clearly distinct density fields (\cref{fig:teaser}(c)), each capturing a different weather pattern for the weather ensemble. 
Unlike point-based techniques~\cite{kumpfVisualizingConfidenceClusterbased2018, zhangEnConVisUnifiedFramework2023}, our method is feature-based and preserves the global coherence of each contour member, since every decoded sample is a complete contour reconstructed through $p(x \mid z)$ rather than an independent per-pixel estimate.

\subsection{Implementation}
\label{sec:implement}
Our method is implemented in Python using \texttt{PyTorch} for the VAE and \texttt{scikit-image} for contour extraction.
The convolutional VAE follows the architecture described in~\cref{sec:vae_method}: the encoder uses three strided convolutional blocks with batch normalization and LeakyReLU (negative slope $0.2$), channel widths $[64, 128, 256]$, and a latent dimension of $k{=}8$; the decoder mirrors this structure with bilinear upsampling followed by convolution, and no activation on the final output layer.
All weights are initialized with Kaiming normal initialization.

We use this compact configuration for all datasets without dataset-specific architecture tuning.
On held-out validation members of the weather, hippocampus, and ventricle ensembles, it yields normalized mean squared errors (NMSEs) of $0.0015$, $0.0117$, and $0.0063$, respectively.
Controlled architecture sensitivity experiments on channel width and latent dimensionality are reported in the supplemental material~\cite{vaeEnVisFrameworkSuppl}.
These results indicate that the common configuration is adequate for the 2D contour ensembles studied here, although $k{=}8$ should be understood as an efficiency-oriented bottleneck rather than an estimate of the exact intrinsic dimensionality or a universally optimal choice.
For a new ensemble, the NMSE on a held-out validation set provides a practical criterion for determining whether additional model capacity is needed.
Substantially more complex or higher-resolution ensembles may require deeper or wider encoders and decoders, or a larger
latent space based on this NMSE-based tuning, but the compact configuration would be a good starting point.

The network is trained for $3{,}000$ epochs with the Adam optimizer~\cite{kingma2017adam} (learning rate $10^{-3}$, batch size~$32$) and a cosine annealing schedule that decays the learning rate to $10^{-5}$.
The KL weight~$\beta$ is linearly annealed from~$0$ to~$2.0$ during the first $1{,}000$ epochs and held constant thereafter; gradient norms are clipped at~$5.0$.

Our method was tested on a workstation with a 3.4~GHz Intel i7 CPU, 32~GB main memory, and an NVIDIA GeForce RTX 4080 Super GPU with 16~GB graphics memory.
For the evaluated datasets, VAE training takes 13.9---65.9~s. Detailed stage-wise timings are reported in the supplemental material~\cite{vaeEnVisFrameworkSuppl}.

\section{Evaluation}
We evaluate our method with numerical comparisons for data depth, clustering, and coherence tests against existing techniques.
Synthetic datasets generated with known distributions (see supplemental material Sec.~7.1 for details) are used to evaluate depth and clustering: an example dataset is shown in~\cref{fig:depthEval}(a).

\subsection{Comparisons of Data Depth}
\label{sec:evalDepth}
We compare our latent space MLS depth to contour band depth (CBD)~\cite{whitakerContourBoxplotsMethod2013} and epsilon inclusion depth (eID)~\cite{chaves-de-plazaInclusionDepthContour2024} on a randomly generated synthetic dataset with 95 ensemble members.
CBD is a well-established contour depth measure, while eID is a computationally efficient approximation of CBD.
\begin{figure}[htb]
\centering
\includegraphics[width=\linewidth]{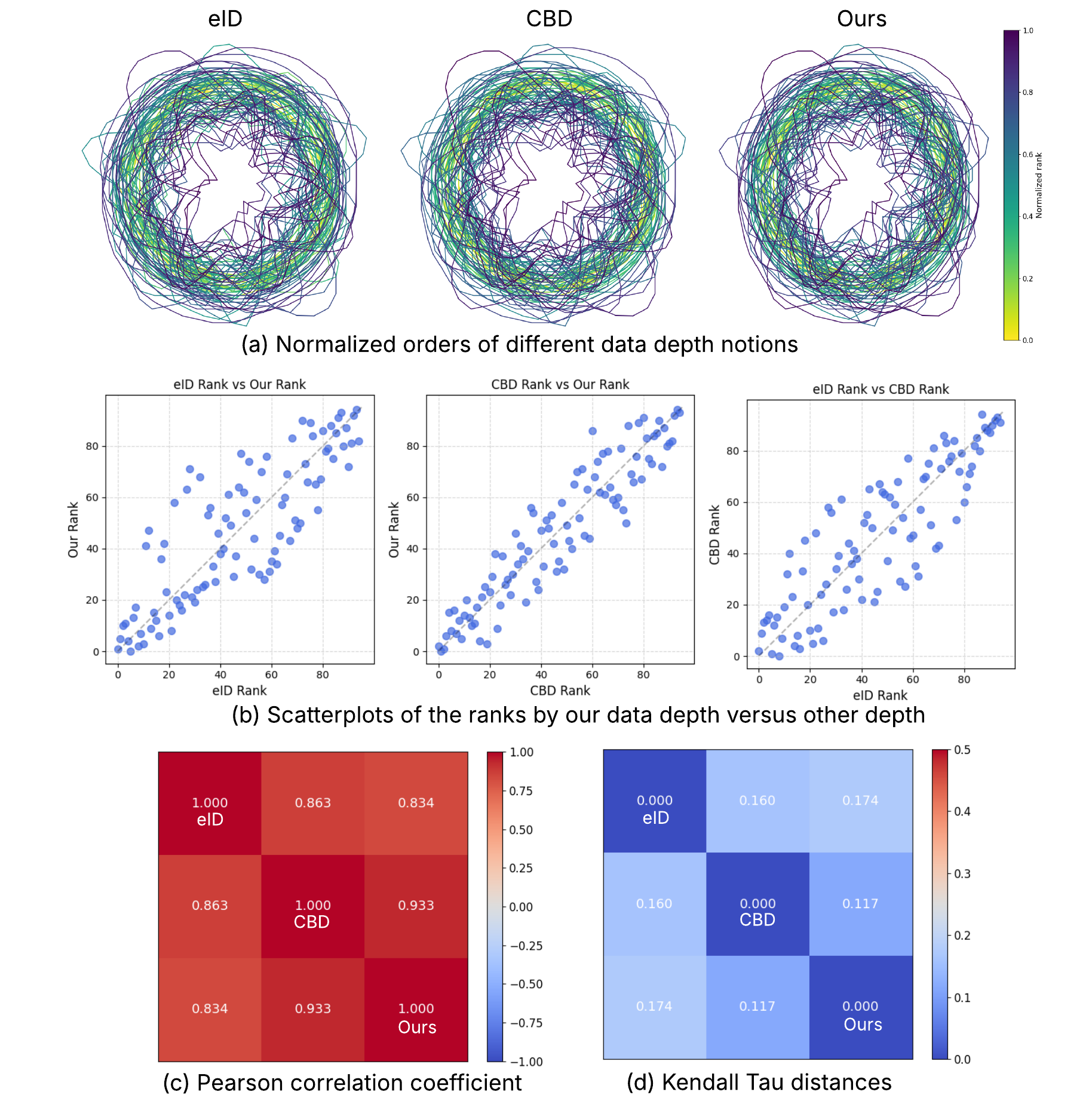}
\caption{Evaluation of data depth comparing our method to CBD and eID.  }
 \label{fig:depthEval}
\end{figure}
Normalized depth orders produced by eID, CBD, and our method are visualized in~\cref{fig:depthEval}(a).
All three methods consistently assign smaller ranks to inner members and larger ranks to outer ones, confirming that our latent space depth captures the same geometric intuition as the contour-based measures.
Pair-wise Pearson correlation coefficients of the resulting orders are shown in~\cref{fig:depthEval}(c).
Our depth is highly correlated with CBD ($0.933$) and moderately correlated with eID ($0.834$), while the correlation between eID and CBD is $0.863$.
The normalized Kendall Tau distance, which quantifies the degree of rank discordancy between two orderings, is reported in~\cref{fig:depthEval}(d).
Our method has a short distance to CBD ($0.117$) and a moderate one to eID ($0.174$); the distance between eID and CBD ($0.160$) is comparable to our distance to eID.
Scatterplots of the pair-wise rank differences in~\cref{fig:depthEval}(b) further confirm this pattern: the scatter between our depth and CBD is tightly concentrated along the diagonal, whereas the spread between our depth and eID is similar to that between CBD and eID.

These results show that our MLS-based depth closely reproduces the ordering of CBD, with the discrepancy to eID being comparable to that between the two contour-based methods themselves.
Crucially, our method achieves this high consistency with a significant computational advantage.
CBD requires computing pairwise band containment over all triples of contours, resulting in $O(N^3)$ complexity, which becomes prohibitive for large ensembles.
In contrast, our depth is derived directly from the MLS matrix, which has $O(N^2)$ complexity and is already precomputed during the latent space construction.
Because the same MLS matrix is also reused for clustering (\cref{sec:evalClustering}) and intra-cluster depth ranking, our depth computation incurs essentially no additional cost beyond what is already required by the pipeline.
Further analysis on cases of large discrepancies in depth comparisons can be found in Sec. 4 of the supplemental material.

\subsection{Clustering Performance}
\label{sec:evalClustering}
Our new clustering method is evaluated by comparing it to existing methods: the standard AHC in our latent space, CD-clustering~\cite{chaves-de-plazaDepthMultimodalContour2024}, and PCA-AHC~\cite{ferstlStreamlineVariabilityPlots2016} on the synthetic radial dataset described in the supplemental material (Sec. 7.1).
These methods are tested on 20 randomly generated ensembles of the synthetic data, where each ensemble contains 50 members generated from two distributions. 
The ground-truth label of each member is given by whether it is generated from the normal or contaminated population.
Because the numerical indices of clustering labels are arbitrary, clustering accuracy is computed as the larger agreement over the two possible assignments between predicted cluster labels and ground-truth labels.
\begin{figure}[htb]
    \centering
    \vspace{-1ex}
    \includegraphics[width=0.8\linewidth]{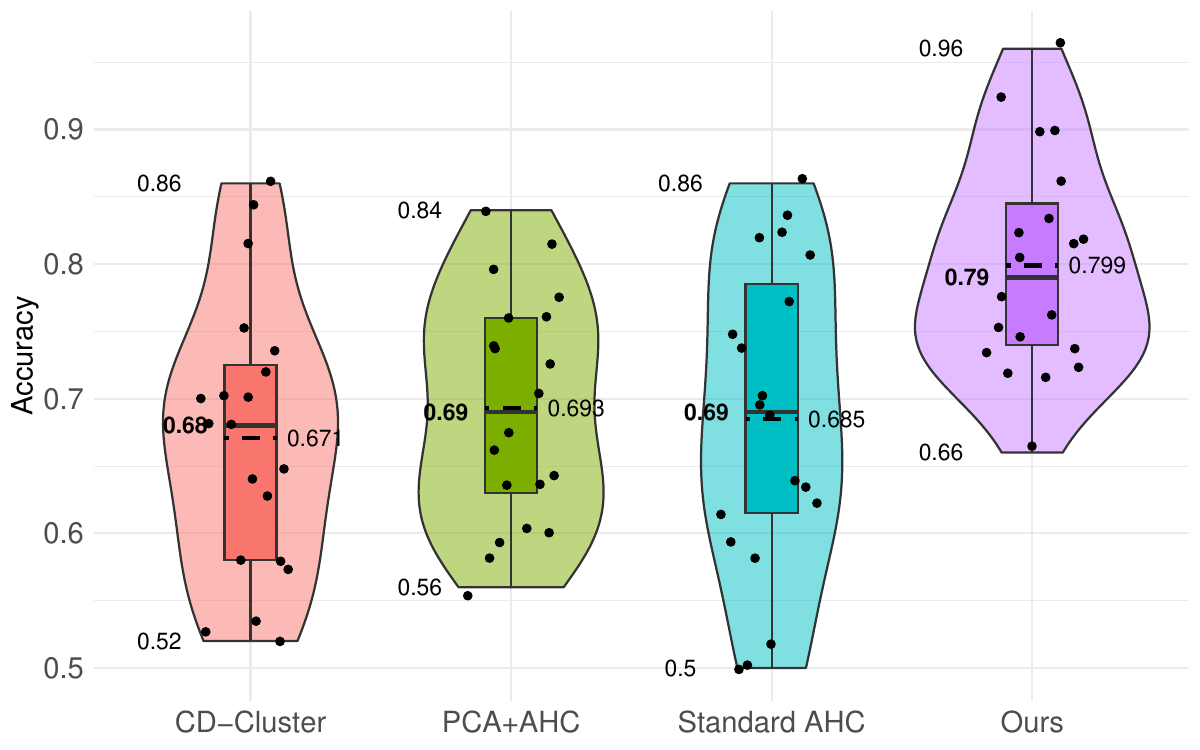}
    \caption{Violin plots and boxplots of accuracies of cluster methods on the synthetic radial dataset. The minimum, maximum, median, and mean (right of boxplots) are labeled for each method.}
    \label{fig:clusterEval}
\end{figure}

Violin plots of clustering accuracy of all techniques are shown in~\cref{fig:clusterEval}.
Our method achieves the highest accuracy compared to PCA-AHC and CD-clustering demonstrates the superior capability in extracting nonlinear features of VAE, enabling it to better capture the subtle distinctions among contour shapes.
Moreover, the improvement over the standard AHC applied in the same latent space indicates that incorporating uncertainty information through MLS-AHC is meaningful, as the probabilistic similarity measure leads to more informed and accurate clustering decisions.

\begin{figure}[htb]
\centering
\subfloat[Ours: cluster 0 \#: 44, cluster \# 1: 51]{\includegraphics[width=0.46\linewidth,trim=0 7.5cm 0 5.2cm,clip]{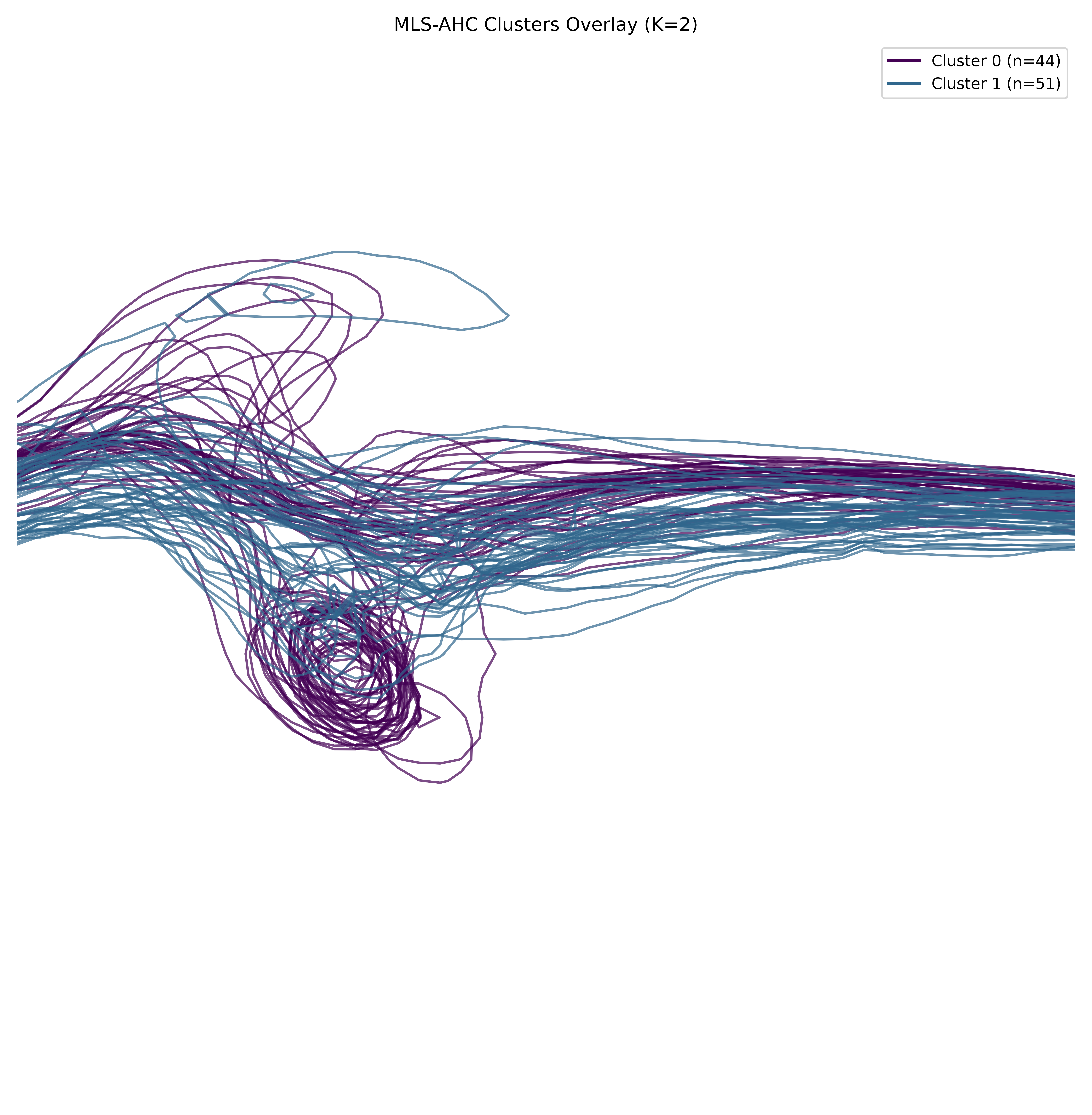}}
\subfloat[PCA-AHC: cluster 0 \#: 10, cluster 1 \#: 85]{\includegraphics[width=0.46\linewidth,trim=0 7.5cm 0 5.2cm,clip]{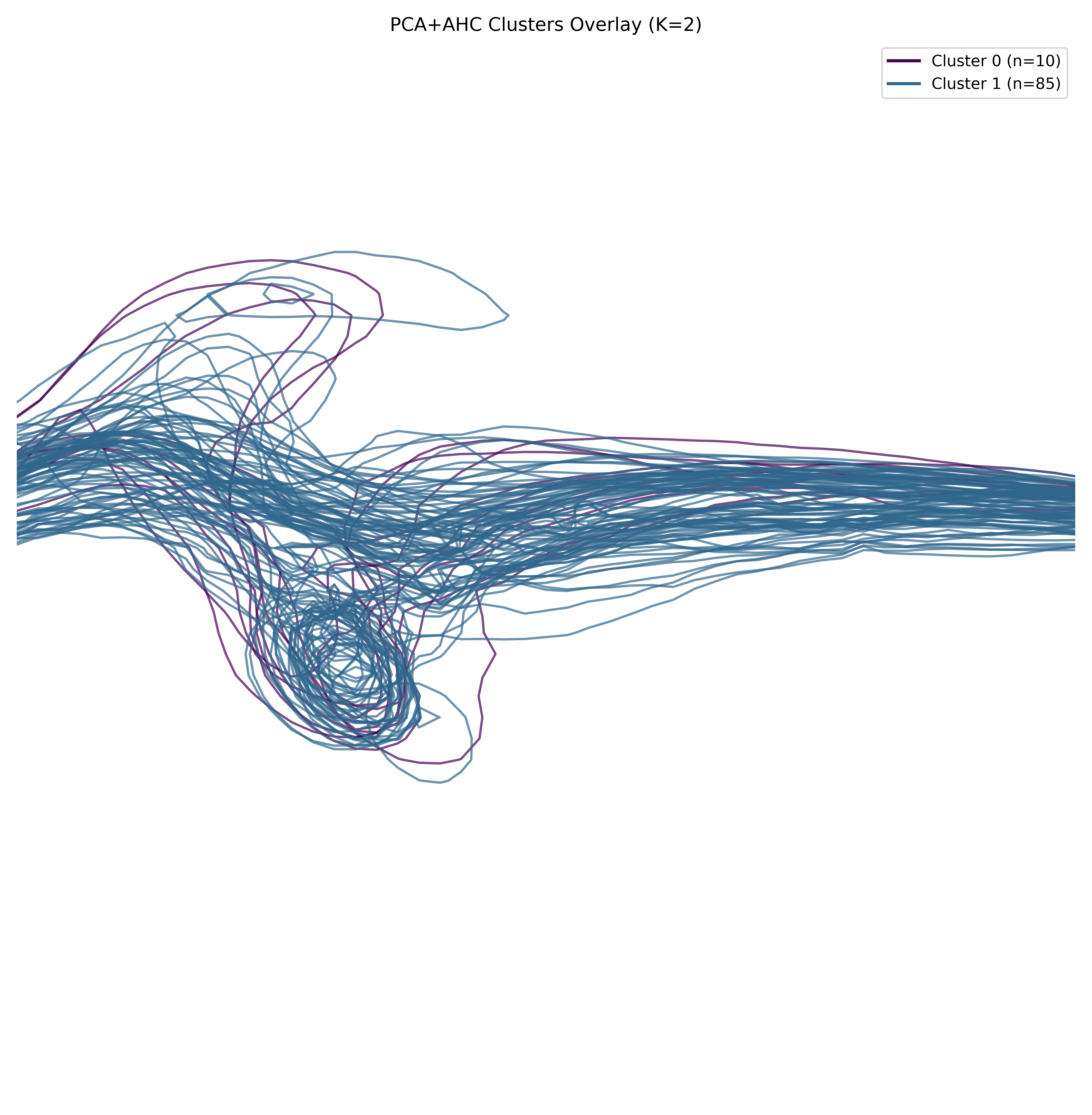}}\\
\subfloat[TTK: cluster 0 \#: 50, cluster 1 \#: 45]{\includegraphics[width=0.46\linewidth,trim=0 7.5cm 0 5.2cm,clip]{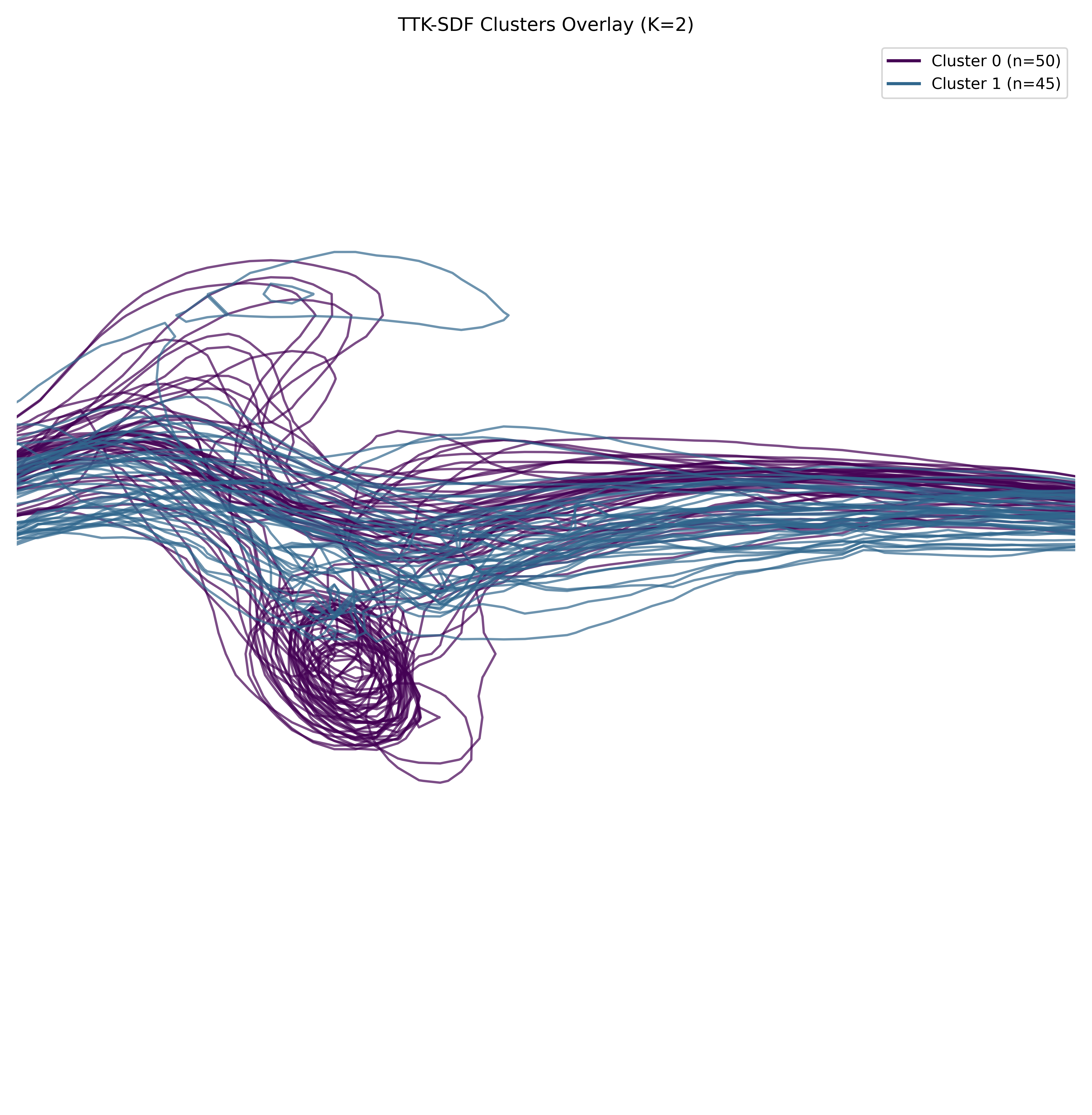}}
\subfloat[CD-clustering: cluster 0 \#: 92, cluster 1 \#: 3]{\includegraphics[width=0.46\linewidth,trim=0 7.5cm 0 5.2cm,clip]{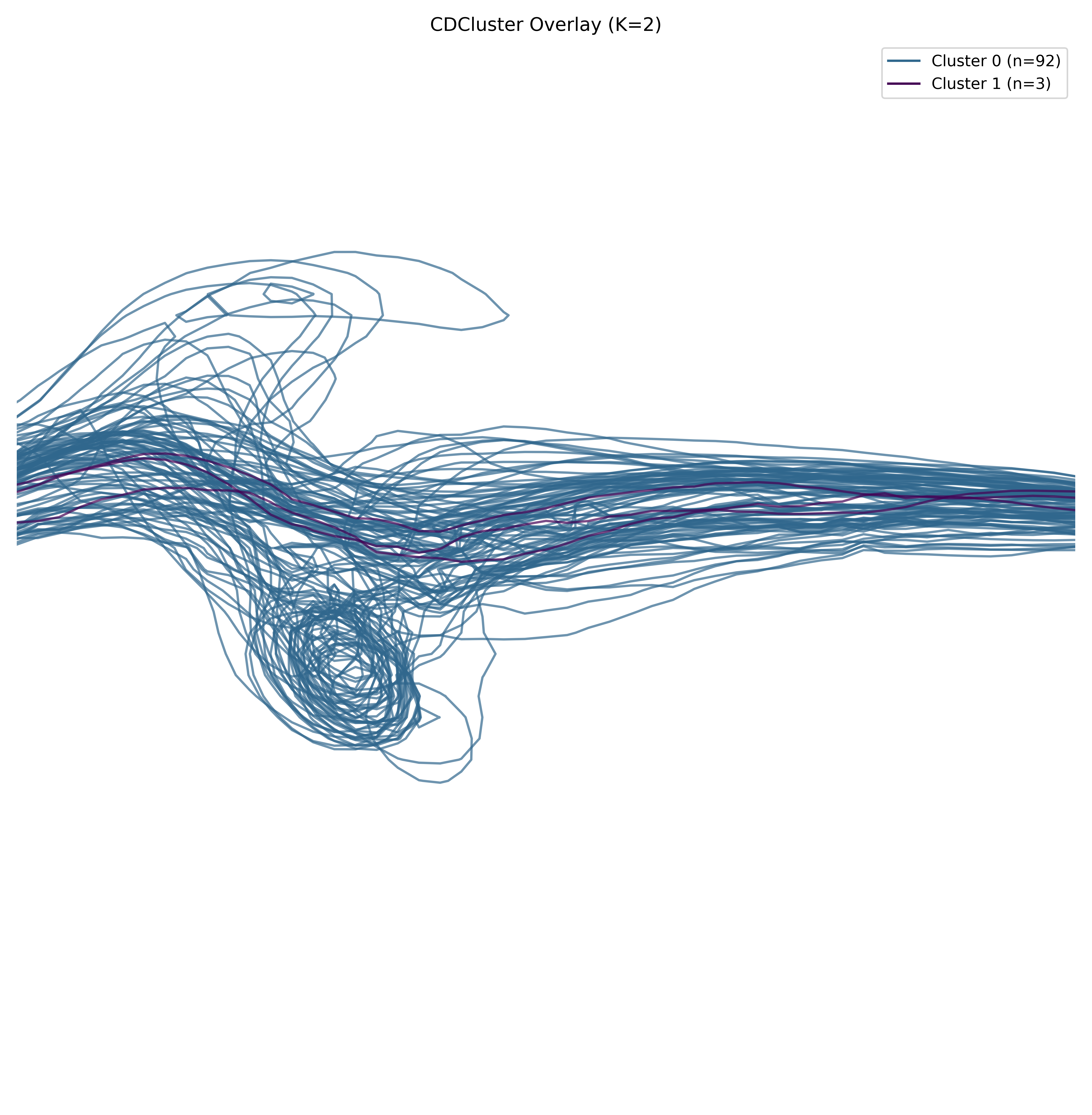}}
 \caption{A comparison of clusters of the weather ensemble generated by (a) our AHC method, (b) PCA-AHC, (c) TTK persistence diagram-based clustering, and (d) CD-clustering. For each method, the number of members in each cluster is reported. }
 \vspace{-1em}
 \label{fig:weatherClusters}
\end{figure}

We further examine whether the clusters reflect topological structures in the weather ensemble (~\cref{fig:weatherClusters}) by comparing our result with the persistence diagram-based clustering method implemented in the Topology ToolKit (TTK)~\cite{Tierny2018,Vidal2020}.
With the number of clusters set to two (blue and purple), our method (\cref{fig:weatherClusters}(a)) and persistence diagram-based clustering (\cref{fig:weatherClusters}(c)) both group members with lower secondary closed-contour structures into one cluster and separate them from those without such structures, which follows the intuition of viewers.
As this real-world case has no ground-truth cluster labels, we report label agreement with TTK instead of clustering accuracy: after label matching, the two partitions agree on 89 of the 95 members, or 93.68\%.
The clusters generated by the two methods contain balanced numbers of members with similar shapes. 
In contrast, PCA-AHC (\cref{fig:weatherClusters}(b)) and CD-clustering (\cref{fig:weatherClusters}(d)) do not recover this structural separation and generate unbalanced partitions.

\subsection{Evaluation of Density Plots}
\label{sec:evalDensity}
We evaluate the probability density estimation of our VAE-based method through qualitative visual comparison and quantitative coherence measurement comparison.
\subsubsection{Visual Quality Comparison}
\begin{figure}[htb]
\centering
\subfloat[Our method]{\includegraphics[width=0.33\linewidth]{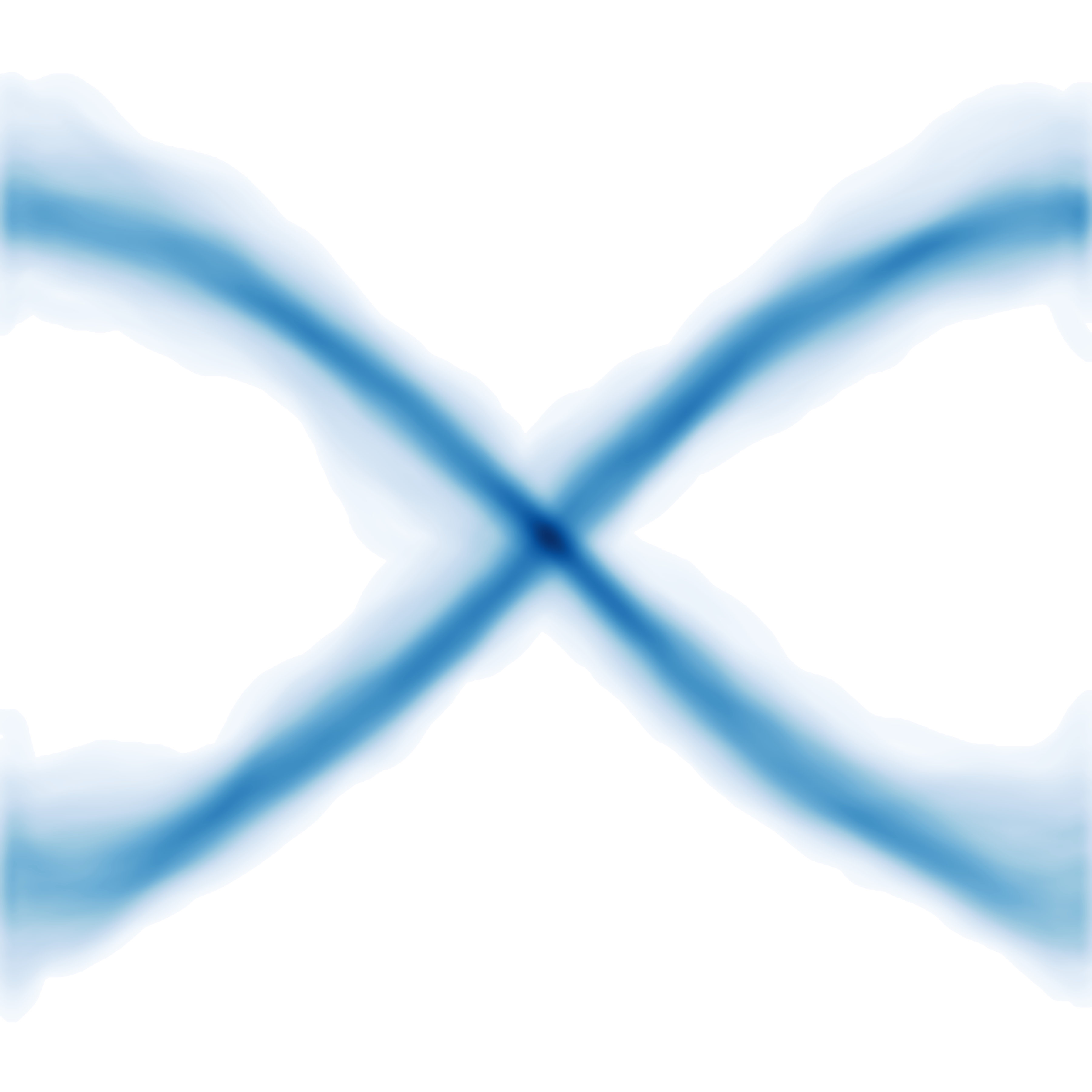}}
% \hspace{1em}
\hfill
\subfloat[EnConVis]{\includegraphics[width=0.33\linewidth]{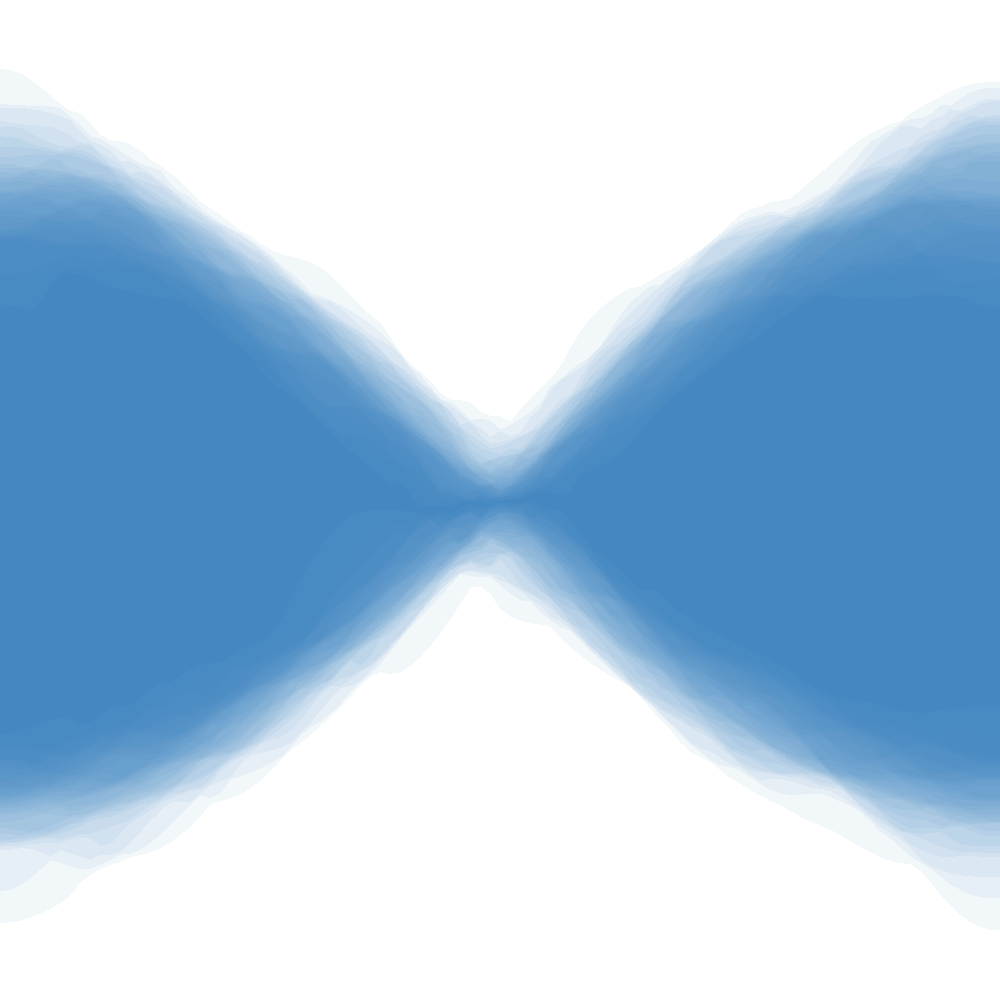}}
\hfill
\subfloat[PDF bands]{\includegraphics[width=0.33\linewidth]{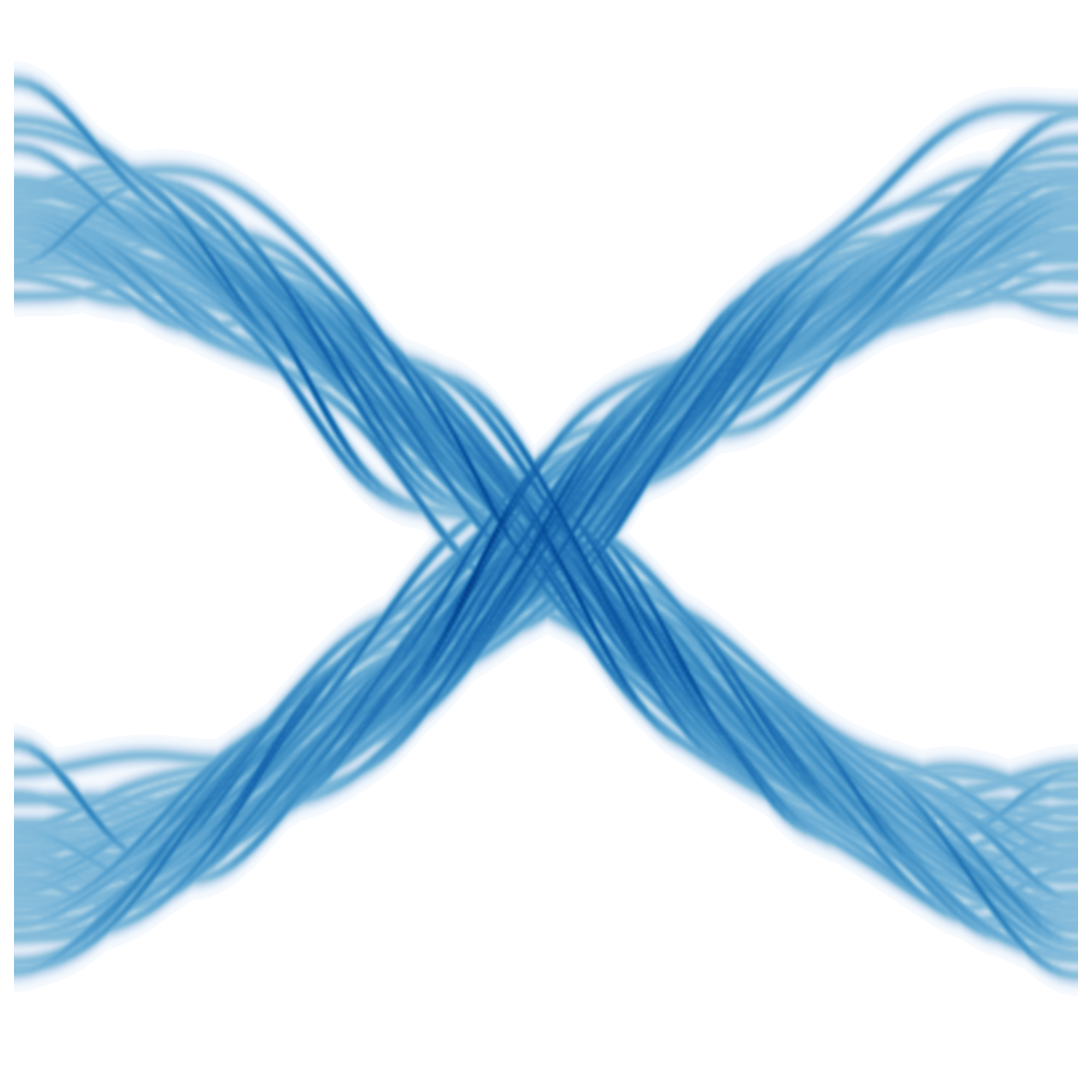}}
\caption{Density estimation on an X-shaped synthetic ensemble of~\cref{fig:clusterProblem}. (a) Our density method correctly recovers the two crossing branches, whereas (b) EnConVis erroneously assigns high probability to the empty region between the two branches. The PDF bands (c) recover ensemble members, but assign high image gradient regions with high probability values. }
\label{fig:densityEval}
\end{figure}

The comparison is performed on the X-shaped synthetic ensemble.
The resulting density plot of our method faithfully recovers the two underlying branches (\cref{fig:densityEval}(a)), with high-density regions coinciding with the actual contour trajectories.
In contrast, EnConVis estimates density at each grid point independently via kernel density estimation, without associating features across spatial locations.
As a result (\cref{fig:densityEval}(b)), EnConVis erroneously assigns high probability to the empty region between the two branches---where no actual contour exists---because contours from both groups contribute to the local kernel on either side, merging both populations into a single misleading high-density band.
Furthermore, we implement the PDF/CDF bands method~\cite{Pfaffelmoser2013}, which smooths the upper/lower iso-region indicator of a member with a Gaussian CDF, differentiates the smoothed field to obtain a spatial PDF, and takes the pointwise maximum across members.
The result of PDF bands (\cref{fig:densityEval}(c)) recovers the two branches and individual members but emphasizes high image gradients with high probabilities.

This comparison shows that our latent-space density estimation, by jointly capturing local and global features, correctly recovers multimodal structure, whereas EnconVis conflates spatially overlapping but structurally distinct populations.
Comparisons of our method and EnConvis can also be seen in~\cref{fig:teaser}(c) and (d).

\subsubsection{Coherency Measurement}
Beyond visual quality, an important advantage of our method is the inherent coherence between its density estimation and depth computation, as both are derived from the same VAE latent space.
In conventional workflows, density and depth are computed by separate, independently developed methods, e.g., kernel density estimation for probability maps~\cite{zhangEnConVisUnifiedFramework2023} and eID or CBD for depth ranking~\cite{chaves-de-plazaInclusionDepthContour2024,whitakerContourBoxplotsMethod2013,wu2025probabilisticinclusiondepthfuzzy}, whose underlying assumptions do not necessarily agree.
This inconsistency may confuse domain experts: a member ranked as deep (central) by the depth measure may reside in a low-density region of the density plot, making the two representations difficult to reconcile.

To quantify, we compute the depth-density coherency coefficient $R^2$ for an ensemble by measuring how well the member ranking by a data depth method aligns with the ordering induced by its own density map.
For every member, we compute the average density accumulated along its contour on the corresponding density plot, then correlate this density-based ordering with the depth ranking.
A high correlation indicates that deeper members consistently pass through higher-density regions---a relationship that is well established in the statistical depth literature~\cite{liu1999multivariate,fraiman1997multivariate,zuo2000general}, since deep members are generated from the most probable part of the distribution.

Our method achieves the highest $R^2$ in all cases, as shown in~\cref{tab:coherency}.
The improvement is smallest for the hippocampus data but larger for the weather, ventricle, and X-shaped ensembles, where more complex contour distributions make independently computed depth and density representations less consistent.
The higher coherency scores indicate that our density plot better explains the depth ordering.
% ---or equivalently, that our depth and density representations are more mutually consistent.
This improved consistency is a consequence of computing both quantities from a shared latent representation, rather than combining independently designed methods that may encode different notions of centrality.

\begin{table}[htb]
    \centering
    \caption{Comparisons of depth--density coherency coefficients $R^2$: higher values indicate better alignment between the depth ranking and the density-based ordering.}
    \label{tab:coherency}
    \begin{tabu}{lccc}
    \toprule
        Dataset & Ours & \shortstack{eID +\\EnConVis~\cite{zhangEnConVisUnifiedFramework2023}} & \shortstack{eID +\\2D PDF~\cite{Pfaffelmoser2013}} \\
        \midrule
        Weather global & 0.677 & 0.370 & 0.372 \\
        Weather cluster 1 & 0.617 & 0.491 & 0.270 \\
        Weather cluster 2 & 0.726 & 0.435 & 0.363 \\
        Hippocampus & 0.752 & 0.732 & 0.669 \\
        Ventricle & 0.931 & 0.818 & 0.800 \\
        Synthetic X-shaped & 0.592 & 0.287 & 0.268 \\
        \bottomrule
    \end{tabu}
\end{table}

\section{Examples}
We demonstrate the usefulness of our method through various real-world ensemble examples from weather forecasting, smoke simulation, and medicine.

\begin{figure}[htb]
    \centering
    \includegraphics[trim={0 1cm 0 1cm},clip,width=0.9\linewidth]{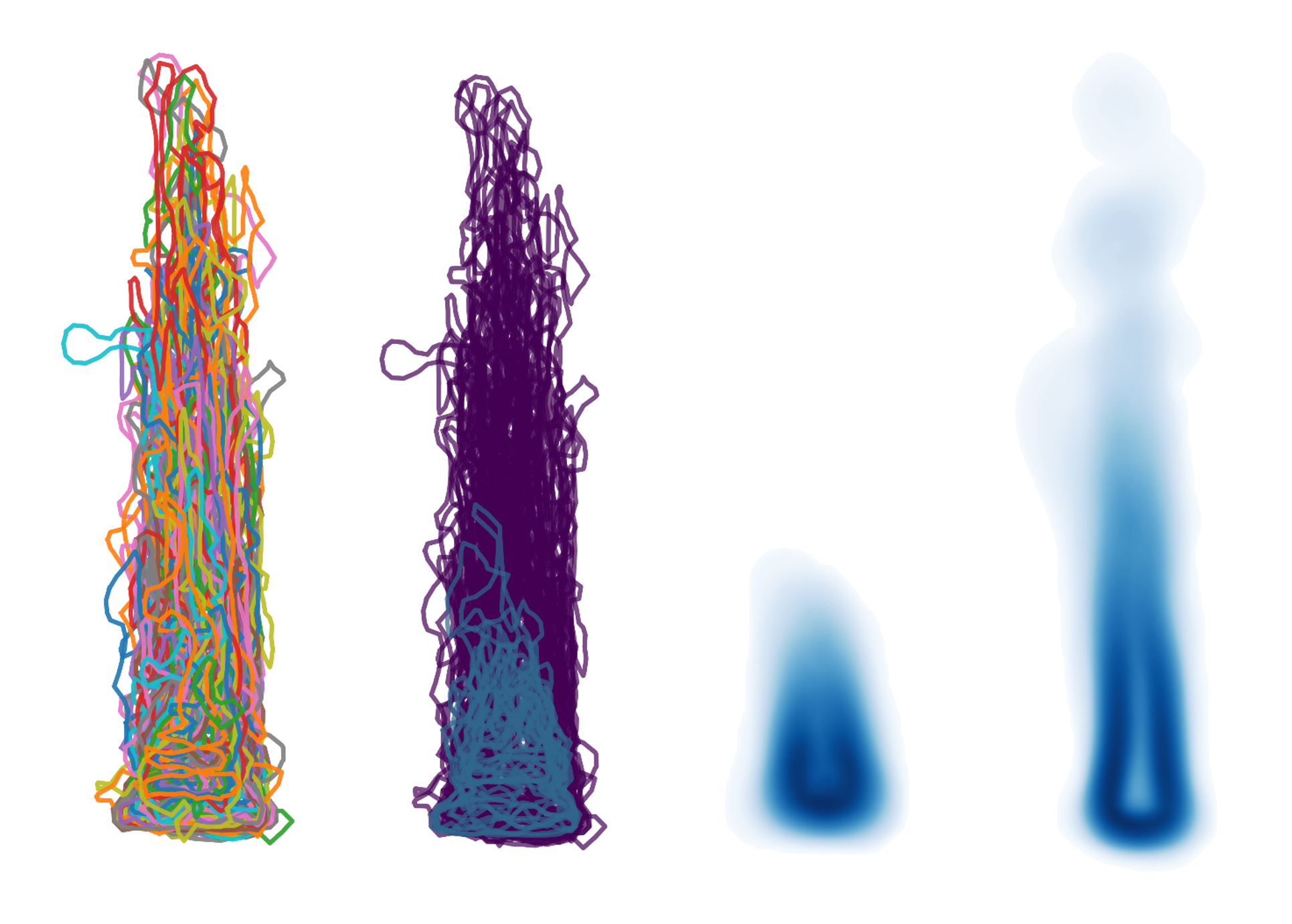}
    \caption{Visualization of 104 smoke plume ensemble members from one slice of ScalarFlow at $t=85$. From left to right: the spaghetti plot of the original isosurface contours, clustered spaghetti plots with two groups (blue and purple), the probability density plots of the blue cluster, and the purple cluster, respectively.}
    \label{fig:scalarflow}
     \vspace{-1ex}
\end{figure}

\subsection{Weather Data}
\label{sec:weather}

The weather dataset is a simulation generated by the ensemble prediction system of the European Centre for Medium-Range Weather Forecasts (ECMWF,~\url{https://www.ecmwf.int}).
We adopt an ensemble of 95 members of a geopotential height field at 500 hPa within the East Asia region~\cite{zhangEnConVisUnifiedFramework2023}.
Results of the dataset are shown in~\cref{fig:teaser}.
The contours form a horizontal main band with a whirling structure at the bottom and a bulge on the top left.
The contours are separated into two groups with our MLS-AHC clustering (\cref{fig:teaser}(b)), whereas the probability density plot visualizes the estimated distributions of the two clusters (\cref{fig:teaser}(c)).
A comparison to the density plot of EnConVis and eID depth for cluster 2 is shown in~\cref{fig:teaser}(d).
\subsection{ScalarFlow}
\label{sec:scalarFlow}

ScalarFlow~\cite{ScalarFlow2019}, a large-scale volumetric dataset of real-world smoke plumes captured with a multi-view setup and reconstructed via a physics-based, simulation-constrained tomography framework, is used as a third example.
We extract isosurface contours from all 104 ensemble members at time step $t=85$ from one slice of the data and visualize them with our method as shown in~\cref{fig:scalarflow}.

The spaghetti plot of the original ensemble (\cref{fig:scalarflow}, leftmost) is heavily cluttered, making it difficult to discern structural patterns from the overlapping contours.
By applying our uncertainty-aware clustering technique, the ensemble is separated into two distinct groups (\cref{fig:scalarflow}, second from left): one cluster (blue) comprises smoke plumes whose mass concentrates at lower elevations, while the other cluster (purple) captures plumes that diffuse and spread upward to higher positions.
The probability density plots of the two clusters (\cref{fig:scalarflow}, right two panels) further reveal the internal structure of each group, delineating where the smoke is more likely to be present within each cluster.

\subsection{Medical Imaging Datasets}
\label{sec:ixidata}
Examples are shown for the publicly available Information eXtraction from Images (IXI) dataset~\cite{antonelli2022medical} and the BRATS2020 brain tumor challenge~\cite{Menze2015,Bakas2017}.
For the IXI dataset, 30 members of 2D slices are used---we take one slice from each member of a total of 30 T1-weighted MRI volumes.
The hippocampus segmentation results using this dataset have been presented in~\cref{fig:haimaDepth}(a).

In addition, our method is applied to the third and fourth ventricles as shown in \cref{fig:haimaDepth}(b).
Compared to the hippocampus (\cref{fig:haimaDepth}(a)), the ventricular contours exhibit a distinctly different variability pattern: their elongated, narrow geometry leads to more pronounced shape variation along the lateral boundaries, as visible in the spaghetti plot (\cref{fig:haimaDepth}(b), middle).
The contour boxplot (\cref{fig:haimaDepth}(b), right) confirms that the overall segmentation uncertainty concentrates along these lateral regions, demonstrating that our depth-based framework generalizes well across anatomical structures with different morphological characteristics.

For the BRATS2020 dataset, the brain segmentation ensembles are generated with a deep-learning segmentation method based on diffusion models~\cite{wollebDiffusionModelsImplicit2022}.
Two ensembles, each with 50 segmentation masks, are calculated for two MRI scan slices of the BRATS2020 challenge.
An example of the tumor segmentation ensemble is shown in~\cref{fig:tumorClusters}, where the variability across segmentation masks reflects the inherent uncertainty in the delineation of tumor boundaries---our MLS-AHC clustering is able to classify them into 4 clusters.

\section{Discussion}
We discuss various aspects of our work, including the neural network architecture, alternative design choices, and limitations.

\paragraph{Neural Network Architecture.}
The common compact architecture (\cref{sec:implement}) achieves low validation NMSEs across the 2D contour ensembles studied in this paper without dataset-specific tuning.
For more complex or higher-resolution data, more expressive architectures can be explored, with reconstruction NMSE on a held-out validation set serving as a practical criterion for assessing their adequacy.

The training of the VAE dominates the one-time initialization cost.
Therefore, when this cost is included, our method is not faster compared to existing methods for a single analysis of an ensemble. 
However, subsequent changes to the number of clusters and within-cluster depth analyses reuse the trained model, encoded member distributions, and complete MLS matrix, avoiding repeated training, encoding, and matrix construction.
A comparison of staged timings to existing methods is documented in the supplemental material.

\paragraph{Alternative Design Choices.}
Other design choices are available for an ensemble visualization framework with latent space exploration.
For example, normalizing flows can model complex non-Gaussian distributions with exact likelihood evaluation and may benefit highly irregular ensembles~\cite{9089305}.
We use a VAE because its regularized low-dimensional latent space is typically easier to train robustly in the small-sample setting, and its member-wise Gaussian posterior directly supports closed-form mutual likelihood scores for data depth and clustering, whereas a standard flow model would require additional definitions of member-wise uncertainty and pairwise similarity.

For the mean contours, optimal transport is available to generate geometrically grounded mean shapes that preserve the topology of ensemble members\cite{7053911}.
We compare the optimal transport with the Wasserstein distance mean shape to our mean contour in Sec. 6 of the supplemental material.
It can be seen that the two shapes look similar overall but are different in detail.
We argue that no ground-truth exists for mean shapes, and no preference can be made there. 
However, a benefit of our mean contour is that it is readily given by the VAE latent space, which is coherent for other tasks of our method.

\paragraph{Limitations.}
Although our neural network architecture is shared across datasets, the VAE is currently retrained for each contour ensemble.
Pretraining on diverse, aligned contour SDFs followed by lightweight fine-tuning may enable transfer to new ensembles, while adaptation-free use remains to be studied.

In theory, our framework is agnostic to the spatial dimension $m$  of the input ensemble.  
However, our current implementation is limited to 2D contour ensembles.  
The extension to 3D ensembles ($m = 3$) or even higher poses computational challenges.
In 3D, 2D SDF images would become volumetric SDFs, and the VAE would need 3D convolutional layers or a more compact neural representation, increasing the cost of encoding, decoding, sampling, and density construction.
The MLS-based depth and clustering steps would still reuse the pairwise similarity matrix, but constructing the feature representation and generating density visualizations would be more expensive in 3D.
Combining our probabilistic latent modeling with implicit neural representations is a promising direction for 3D extensions: a coordinate-based decoder could query 3D SDF values without generating a full volumetric grid for every latent sample, reducing density-generation costs while preserving the latent-space basis for depth and clustering.

\section{Conclusion}
We have presented a probabilistic latent space modeling method for coherent 2D contour ensemble visualization.
By modeling ensemble members as multidimensional distributions in the nonlinear latent space of a VAE, we support various important tasks of ensemble data analysis, including data depth computation and ensemble member clustering.
These tasks are performed efficiently with a similarity matrix that describes pair-wise similarities between members. 
Using the nonlinear transformation from the latent space to the spatial domain of the VAE, our method generates density plots with global coherency information.  
Our method is evaluated by ranking consistency tests for data depth, accuracy tests for clustering, and coherence tests for the alignment of density plots and data depth.
Compared to existing techniques, our method can correctly handle multimodal distributions and offer more coherency between different visual representations of ensemble members.

For future work, an important direction is to extend the framework to 3D ensembles.
For spatiotemporal ensembles, we plan to explore time-consistent latent representations that support feature tracking and visual analysis of temporal evolution.
Using flow-based models, diffusion~\cite{Ho2020,Rombach2022}, or StyleGAN~\cite{Karras2019} techniques for more expressive density modeling and high-fidelity ensemble member generation is also of interest.
Finally, we would like to conduct a formal user study to evaluate the benefits of the improved coherence of our method in data understanding and decision making.

% %% if specified like this the section will be omitted in review mode
\acknowledgments{%
   This work was supported in part by the National Science Foundation of China (grant no.~62372012), and the Beijing Natural Science Foundation (grant no.~4262025).
}

\bibliographystyle{abbrv-doi-hyperref}

\bibliography{datadepth}

\appendix % You can use the `hideappendix` class option to skip everything after \appendix
\crefalias{section}{appendix} % this is to make sure that cleverref switches to referring to Appx. X from here on

\section{Organization of the Appendix}
This appendix covers contents that are left out from our paper for conciseness.
We organize the content as follows.
\begin{itemize}
    \item \cref{sec:mls_derivation} provides the complete mathematical derivation of the Mutual Likelihood Score (MLS), the probabilistic similarity measure that underpins the precomputed pairwise similarity matrix $\mathbf{M}$ used throughout the framework.
    \item \cref{sec:depth_equivalence} formally proves that the Mahalanobis depth reduces to the Euclidean depth under isotropic latent spaces, and that the Euclidean depth ranking is equivalent to a maximum likelihood ranking, thereby motivating our generalization to MLS-based probabilistic depth.
    \item \cref{sec:depth_cases} examines the ensemble members with the largest depth-ranking discrepancies among our MLS depth, contour band depth (CBD), and epsilon inclusion depth (eID), revealing that the discrepancies are primarily driven by eID's local inclusion criterion.
    \item \cref{sec:cvp_comparison} compares our VAE-based density estimation with the PCA-based contour variational plot (CVP), demonstrating the advantages of nonlinear feature extraction and structured latent sampling for probability density plots.
    \item \cref{sec:representative_contour} presents representative contours decoded from the latent space centroid $\bar{\mu}$ for medical and weather ensembles, illustrating how the VAE decoder produces geometrically coherent summary shapes.
    \item \cref{sec:synthetic_data} details the generation procedures for the radial and cross-shaped synthetic datasets used in our evaluations.
    \item \rev{\cref{sec:network_configuration} evaluates the common VAE configuration through controlled architecture sensitivity experiments on held-out validation members and discusses practical model-adequacy diagnostics.}
    \item \rev{\cref{sec:measured_runtime} reports measured stage-wise running times, including VAE training and comparisons with conventional clustering, depth, and density pipelines.}

    % \item \cref{sec:large_figures} provides enlarged versions of all main-text figures for improved readability.
\end{itemize}

\section{Derivation of the mutual likelihood score}
\label{sec:mls_derivation}

This section provides the complete mathematical derivation of the Mutual Likelihood Score (MLS) formula (Eq.~6 in the paper) from the coincidence probability definition (Eq.~5 in the paper).

\subsection{Problem Setup}
As described in the paper, the VAE encoder maps each ensemble member $x_i$ to a diagonal Gaussian posterior in the $k$-dimensional latent space $Z$:
\begin{equation}
    p(z \mid x_i)
    \approx \mathcal{N}\!\left(\mu_i,\,\mathrm{diag}(\sigma_i^2)\right)
    = \prod_{l=1}^{k}
      \mathcal{N}\!\left(z(l);\,\mu_i(l),\,\sigma_i^2(l)\right),
\end{equation}
where $\mu_i(l)$ and $\sigma_i^2(l)$ denote the mean and variance of the $l$-th latent dimension, respectively.

Given two members $x_i$ and $x_j$ with latent representations $z_i \sim p(z \mid x_i)$ and $z_j \sim p(z \mid x_j)$, we seek a closed-form expression for their similarity defined by the probability that $z_i$ and $z_j$ coincide.

\subsection{Derivation}

\noindent\textbf{Step 1: Formulating the coincidence probability.}
The probability that two independently sampled latent variables $z_i$ and $z_j$ take the same value is expressed using the Dirac delta function $\delta(\cdot)$:
\begin{equation}
    p(z_i = z_j)
    = \int \!\!\int p(z_i \mid x_i)\, p(z_j \mid x_j)\,
      \delta(z_i - z_j)\, dz_i\, dz_j\;.
    \label{eqn:supp_pzij}
\end{equation}

\noindent\textbf{Step 2: Reducing to an overlap integral.}
Integrating over $z_j$ first using the sifting property $\int f(z_j)\,\delta(z_i - z_j)\,dz_j = f(z_i)$ yields:
\begin{equation}
    p(z_i = z_j) = \int p(z \mid x_i)\, p(z \mid x_j)\, dz\;,
    \label{eqn:supp_overlap}
\end{equation}
where we rename the remaining integration variable to $z$ for clarity.
This is the overlap integral of two multivariate Gaussian densities.
Since the covariance matrices are diagonal, the integral factorizes over the $k$ latent dimensions:
\begin{equation}
    p(z_i = z_j)
    = \prod_{l=1}^{k} \int_{-\infty}^{\infty}
      \mathcal{N}\!\left(z;\,\mu_i(l),\,\sigma_i^2(l)\right)
      \cdot
      \mathcal{N}\!\left(z;\,\mu_j(l),\,\sigma_j^2(l)\right) dz\;.
    \label{eqn:supp_factorize}
\end{equation}
It therefore suffices to evaluate the one-dimensional integral for each dimension~$l$.

\noindent\textbf{Step 3: Evaluating the one-dimensional overlap integral.}
For a single dimension~$l$, substituting the Gaussian density
$\mathcal{N}(z;\,\mu,\,\sigma^2)
 = \frac{1}{\sqrt{2\pi}\,\sigma}
   \exp\!\bigl(-\frac{(z-\mu)^2}{2\sigma^2}\bigr)$,
the product of the two densities is:
\begin{align}
    &\mathcal{N}\!\left(z;\,\mu_i,\,\sigma_i^2\right)
     \cdot
     \mathcal{N}\!\left(z;\,\mu_j,\,\sigma_j^2\right)
    \nonumber\\
    &= \frac{1}{2\pi\,\sigma_i\,\sigma_j}
       \exp\!\left(
         -\frac{(z-\mu_i)^2}{2\sigma_i^2}
         -\frac{(z-\mu_j)^2}{2\sigma_j^2}
       \right),
    \label{eqn:supp_product}
\end{align}
where we drop the index~$(l)$ for brevity in this step.

We combine the two quadratic terms in the exponent by completing the square with respect to~$z$.
Define the combined variance and mean:
\begin{equation}
    \sigma_c^2
    = \frac{\sigma_i^2\,\sigma_j^2}{\sigma_i^2 + \sigma_j^2}\;,
    \qquad
    \mu_c
    = \frac{\mu_i\,\sigma_j^2 + \mu_j\,\sigma_i^2}
           {\sigma_i^2 + \sigma_j^2}\;.
    \label{eqn:supp_combined}
\end{equation}
Then the sum of the two quadratic forms decomposes as:
\begin{align}
    \frac{(z-\mu_i)^2}{\sigma_i^2}
    + \frac{(z-\mu_j)^2}{\sigma_j^2}
    = \frac{(z-\mu_c)^2}{\sigma_c^2}
    + \frac{(\mu_i-\mu_j)^2}{\sigma_i^2+\sigma_j^2}\;,
    \label{eqn:supp_complete_square}
\end{align}
where the first term on the right depends on~$z$ and defines a Gaussian kernel, and the second term is constant with respect to~$z$.

Substituting~(\ref{eqn:supp_complete_square}) into~(\ref{eqn:supp_product}), the integrand separates into a $z$-dependent Gaussian kernel and a constant factor:
\begin{align}
    &\int_{-\infty}^{\infty}
      \mathcal{N}\!\left(z;\,\mu_i,\,\sigma_i^2\right)
      \cdot
      \mathcal{N}\!\left(z;\,\mu_j,\,\sigma_j^2\right) dz
    \nonumber\\
    &= \frac{1}{2\pi\,\sigma_i\,\sigma_j}\,
       \exp\!\left(
         -\frac{(\mu_i-\mu_j)^2}{2(\sigma_i^2+\sigma_j^2)}
       \right)
    \nonumber\\
    &\quad\;\cdot
       \int_{-\infty}^{\infty}
         \exp\!\left(
           -\frac{(z-\mu_c)^2}{2\sigma_c^2}
         \right) dz\;.
    \label{eqn:supp_integral_step}
\end{align}
By the standard Gaussian integral formula, the remaining integral evaluates to $\sqrt{2\pi}\,\sigma_c$.

Substituting $\sigma_c = \sigma_i\,\sigma_j\big/\!\sqrt{\sigma_i^2+\sigma_j^2}$ from~(\ref{eqn:supp_combined}) and simplifying:
\begin{equation}
    \frac{1}{2\pi\,\sigma_i\,\sigma_j}
    \cdot \sqrt{2\pi}
    \cdot \frac{\sigma_i\,\sigma_j}{\sqrt{\sigma_i^2+\sigma_j^2}}
    = \frac{1}{\sqrt{2\pi\!\left(\sigma_i^2+\sigma_j^2\right)}}\;.
\end{equation}

Therefore, restoring the dimension index~$(l)$, the per-dimension overlap integral is:
\begin{align}
    &\int_{-\infty}^{\infty}
      \mathcal{N}\!\left(z;\,\mu_i(l),\,\sigma_i^2(l)\right)
      \cdot
      \mathcal{N}\!\left(z;\,\mu_j(l),\,\sigma_j^2(l)\right) dz
    \nonumber\\
    &= \frac{1}{\sqrt{2\pi\!\left(\sigma_i^2(l)+\sigma_j^2(l)\right)}}\,
       \exp\!\left(
         -\frac{(\mu_i(l)-\mu_j(l))^2}
              {2\left(\sigma_i^2(l)+\sigma_j^2(l)\right)}
       \right).
    \label{eqn:supp_perdim}
\end{align}

\noindent\textbf{Step 4: Combining all $k$ dimensions.}
Substituting~(\ref{eqn:supp_perdim}) back into the factorized form~(\ref{eqn:supp_factorize}):
\begin{align}
    p(z_i = z_j)
    &= \prod_{l=1}^{k}
       \frac{1}{\sqrt{2\pi\!\left(\sigma_i^2(l)+\sigma_j^2(l)\right)}}
    \nonumber\\
    &\qquad\;\cdot\,
       \exp\!\left(
         -\frac{(\mu_i(l)-\mu_j(l))^2}
              {2\left(\sigma_i^2(l)+\sigma_j^2(l)\right)}
       \right).
    \label{eqn:supp_full}
\end{align}

\noindent\textbf{Step 5: Taking the logarithm to obtain the MLS.}
The MLS is defined as the log-likelihood of the coincidence probability.
Taking the logarithm of~(\ref{eqn:supp_full}) and converting the product into a sum:
\begin{align}
    &\text{MLS}(z_i, z_j)
    = \log p(z_i = z_j)
    \nonumber\\
    &= \sum_{l=1}^{k} \biggl[
         -\frac{1}{2}\log 2\pi
         - \frac{1}{2}\log\!\left(\sigma_i^2(l)+\sigma_j^2(l)\right)
    \nonumber\\
    &\qquad\qquad
         - \frac{(\mu_i(l)-\mu_j(l))^2}
              {2\left(\sigma_i^2(l)+\sigma_j^2(l)\right)}
       \biggr]
    \nonumber\\
    &= -\frac{1}{2}\sum_{l=1}^{k} \left(
         \frac{(\mu_i(l)-\mu_j(l))^2}{\sigma_i^2(l)+\sigma_j^2(l)}
         + \log\!\left(\sigma_i^2(l)+\sigma_j^2(l)\right)
       \right)
    \nonumber\\
    &\quad\; - \frac{k}{2}\log 2\pi\;.
    \label{eqn:supp_mls_final}
\end{align}

Letting $C = \frac{k}{2}\log 2\pi$, we arrive at the closed-form MLS as stated in Eq.~6 of the paper:
\begin{equation}
  \text{MLS} = -\frac{1}{2}\sum_{l=1}^{k} \left(
    \frac{(\mu_i(l)-\mu_j(l))^2}{\sigma_i^2(l)+\sigma_j^2(l)}
    + \log\left(\sigma_i^2(l)+\sigma_j^2(l)\right)
  \right) - C\;.
\end{equation}

\subsection{Interpretation}
The MLS consists of two terms summed over all $k$ latent dimensions:

\begin{itemize}
    \item \textbf{Squared Mahalanobis-like term}
    $\displaystyle\frac{(\mu_i(l)-\mu_j(l))^2}{\sigma_i^2(l)+\sigma_j^2(l)}$:
    measures the squared difference of the means normalized by the sum of the variances.
    This term penalizes members whose latent means are far apart, with the penalty reduced when either member has high encoding uncertainty.

    \item \textbf{Log-variance term}
    $\displaystyle\log\!\left(\sigma_i^2(l)+\sigma_j^2(l)\right)$:
    penalizes pairs with large combined uncertainty.
    Even if two members have identical means, large variances reduce the MLS because their distributions are more diffuse, lowering the peak of the overlap density.
\end{itemize}

The constant $C = \frac{k}{2}\log 2\pi$ is a normalization term independent of the member pair, and thus does not affect relative rankings.
The MLS is symmetric, i.e., $\text{MLS}(z_i, z_j) = \text{MLS}(z_j, z_i)$, as every term is symmetric in $i$ and $j$.
This symmetry ensures that the resulting similarity matrix $\mathbf{M}$ is symmetric, as required by the downstream depth and clustering algorithms.

\section{Depth equivalence under isotropy}
\label{sec:depth_equivalence}

This section provides the formal justification for two claims made in Section~3.3 of the paper:
(1)~the Mahalanobis depth reduces to the Euclidean depth in the isotropic latent space of the VAE, and
(2)~minimizing the average pairwise squared Euclidean distance is mathematically equivalent to maximizing the joint log-likelihood under an isotropic Gaussian assumption.

\subsection{From Mahalanobis Depth to Euclidean Depth}
The Mahalanobis depth of a member $z_i$ with respect to a distribution with mean $\mu$ and covariance $\Sigma$ is defined as:
\begin{equation}
    D_M(z_i) = \frac{1}{1 + (z_i-\mu)^\top \Sigma^{-1} (z_i-\mu)}\;.
    \label{eqn:supp_mahal_depth}
\end{equation}

The VAE is trained with a standard normal prior $p(z) = \mathcal{N}(0, I)$, and the KL regularization term in the training objective encourages the aggregate posterior to approximate this prior.
Consequently, the latent space is approximately isotropic, and the empirical covariance of the encoded members is well approximated by $\Sigma \approx \sigma^2 I$ for some scalar $\sigma^2 > 0$.

Under this isotropic assumption, $\Sigma^{-1} = \frac{1}{\sigma^2} I$, and the quadratic form in~(\ref{eqn:supp_mahal_depth}) simplifies to:
\begin{equation}
    (z_i-\mu)^\top \Sigma^{-1} (z_i-\mu)
    = \frac{1}{\sigma^2} (z_i-\mu)^\top (z_i-\mu)
    = \frac{\|z_i - \mu\|_2^2}{\sigma^2}\;.
    \label{eqn:supp_mahal_simplify}
\end{equation}

Therefore, the Mahalanobis depth becomes:
\begin{equation}
    D_M(z_i) = \frac{1}{1 + \|z_i - \mu\|_2^2 / \sigma^2}\;,
    \label{eqn:supp_mahal_euclidean}
\end{equation}
which is a strictly decreasing function of the squared Euclidean distance $\|z_i - \mu\|_2^2$.
Since $\sigma^2$ is a shared constant, the depth ranking induced by $D_M$ is identical to that induced by the Euclidean distance to $\mu$.

In practice, the distribution mean $\mu$ is estimated by the sample mean $\bar{z} = \frac{1}{N}\sum_{j=1}^{N} z_j$.
It is a well-known identity that the sum of squared distances from any point to all data points decomposes as:
\begin{equation}
    \sum_{j=1}^{N} \|z_i - z_j\|_2^2
    = N\|z_i - \bar{z}\|_2^2 + \sum_{j=1}^{N} \|\bar{z} - z_j\|_2^2\;.
    \label{eqn:supp_variance_decomp}
\end{equation}
Since the second term on the right is independent of $z_i$, ranking members by $\|z_i - \bar{z}\|_2^2$ is equivalent to ranking them by the average pairwise squared distance $\frac{1}{N}\sum_{j=1}^{N}\|z_i - z_j\|_2^2$.
This yields the Euclidean depth:
\begin{equation}
    D_E(z_i) \propto \left( \frac{1}{N} \sum_{j=1}^{N} \|z_i - z_j\|_2^2 \right)^{-1},
    \label{eqn:supp_euclidean_depth}
\end{equation}
which produces the same member ranking as the Mahalanobis depth under isotropy.

\subsection{Euclidean Depth as Maximum Likelihood Depth}
We now show that the Euclidean depth ranking coincides with the ranking by joint log-likelihood under the isotropic Gaussian model, thereby providing a probabilistic interpretation of the geometric depth.

\noindent\textbf{Step 1: Isotropic Gaussian likelihood.}
Assume that each member $z_j$ is independently drawn from an isotropic Gaussian centered at a candidate center $z_i$:
\begin{equation}
    p(z_j \mid z_i) = \mathcal{N}(z_j;\, z_i,\, \sigma^2 I)
    = \frac{1}{(2\pi\sigma^2)^{k/2}}
      \exp\!\left( -\frac{\|z_j - z_i\|_2^2}{2\sigma^2} \right).
    \label{eqn:supp_isotropic}
\end{equation}

\noindent\textbf{Step 2: Log-likelihood of a single observation.}
Taking the logarithm of~(\ref{eqn:supp_isotropic}):
\begin{equation}
    \log p(z_j \mid z_i)
    = -\frac{k}{2}\log(2\pi\sigma^2)
       - \frac{\|z_j - z_i\|_2^2}{2\sigma^2}\;.
    \label{eqn:supp_single_loglik}
\end{equation}
The first term is a constant independent of the pair $(z_i, z_j)$.
Therefore, the log-likelihood is a strictly decreasing affine function of $\|z_j - z_i\|_2^2$.

\noindent\textbf{Step 3: Joint log-likelihood over the ensemble.}
Summing over all $N$ members:
\begin{align}
    \mathcal{L}(z_i)
    &= \sum_{j=1}^{N} \log p(z_j \mid z_i)
    \nonumber\\
    &= -\frac{Nk}{2}\log(2\pi\sigma^2)
       - \frac{1}{2\sigma^2}\sum_{j=1}^{N} \|z_j - z_i\|_2^2\;.
    \label{eqn:supp_joint_loglik}
\end{align}

\noindent\textbf{Step 4: Establishing the equivalence.}
Since $\sigma^2 > 0$ and $N$, $k$ are fixed, the first term in~(\ref{eqn:supp_joint_loglik}) is constant with respect to $z_i$.
Maximizing the joint log-likelihood therefore reduces to minimizing the sum of squared distances:
\begin{equation}
    \arg\max_{z_i}\; \mathcal{L}(z_i)
    = \arg\min_{z_i}\; \sum_{j=1}^{N} \|z_j - z_i\|_2^2\;.
    \label{eqn:supp_equivalence}
\end{equation}
This is identical to the Euclidean depth criterion in~(\ref{eqn:supp_euclidean_depth}).

More generally, the complete ranking is preserved.
For any two members $z_a$ and $z_b$:
\begin{align}
    D_E(z_a) \geq D_E(z_b)
    &\;\Longleftrightarrow\;
    \sum_{j=1}^{N} \|z_a - z_j\|_2^2
    \leq
    \sum_{j=1}^{N} \|z_b - z_j\|_2^2
    \nonumber\\
    &\;\Longleftrightarrow\;
    \mathcal{L}(z_a) \geq \mathcal{L}(z_b)\;,
    \label{eqn:supp_rank_equiv}
\end{align}
where the second equivalence follows directly from~(\ref{eqn:supp_joint_loglik}).

\subsection{Remark}
The above equivalences rely on the isotropic assumption $\Sigma = \sigma^2 I$.
When each ensemble member is represented not as a deterministic point but as a distribution with its own variance---as in our VAE framework---the isotropic assumption no longer captures the full probabilistic structure.
This limitation motivates the generalization to MLS-based depth presented in Section~3.3 of the paper, where the per-member variance is explicitly incorporated into the similarity and depth computation.

\section{Cases in Depth Comparison with Large Discrepancies}
\label{sec:depth_cases}
\begin{figure}[htb]
\centering
\includegraphics[width=0.9\linewidth]{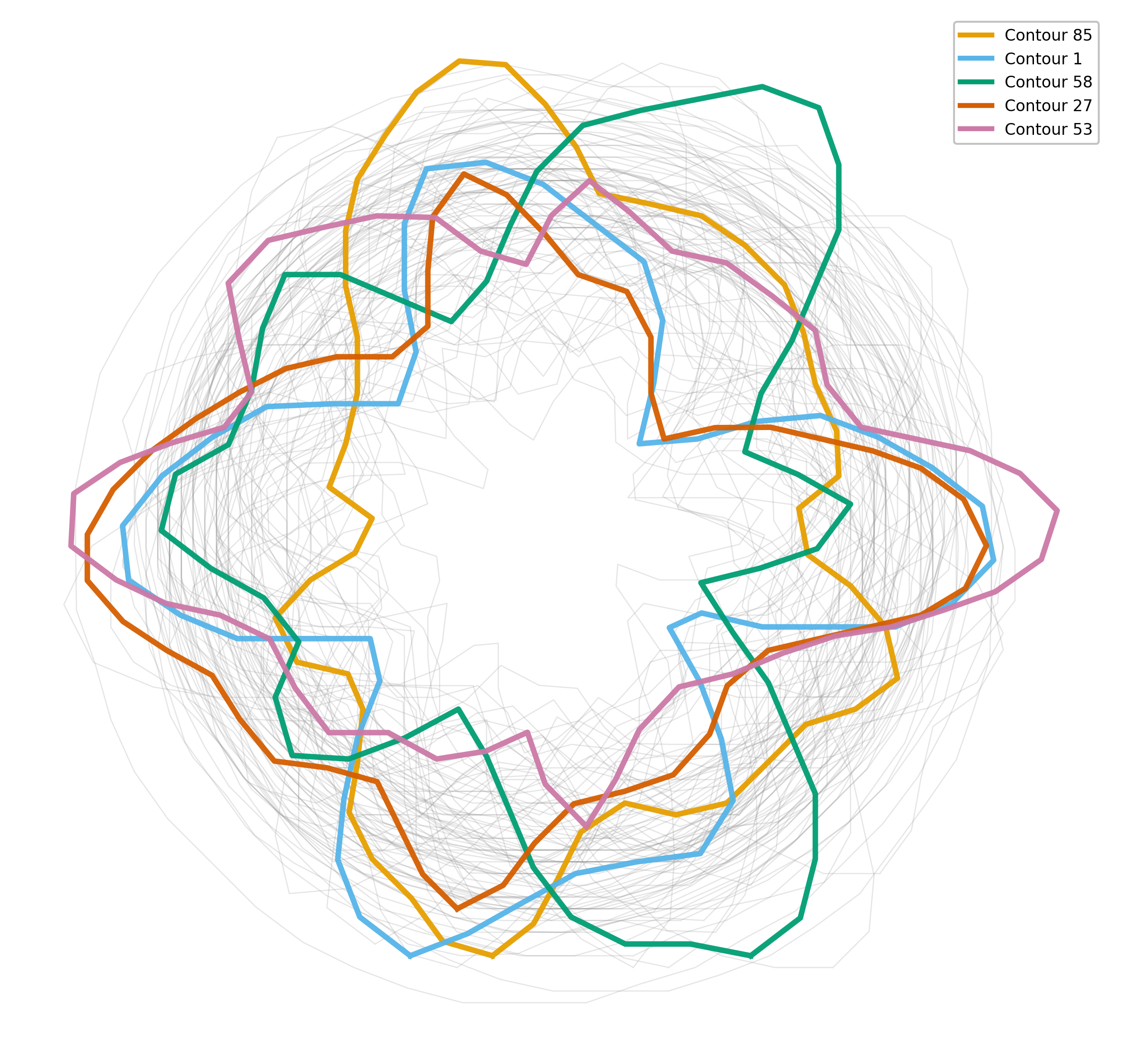}
\caption{The top five ensemble member contours with the largest ranking differences among our MLS depth, eID, and CBD. Despite overall high correlation, these cases highlight where the three methods diverge most in their depth ordering.}
\label{fig:worstCases}
\end{figure}

\begin{figure*}[tb]
    \centering
    \subfloat[Probability density plot based on PCA-CVP~\cite{zhangEnConVisUnifiedFramework2023}]{%
        \includegraphics[width=0.33\textwidth]{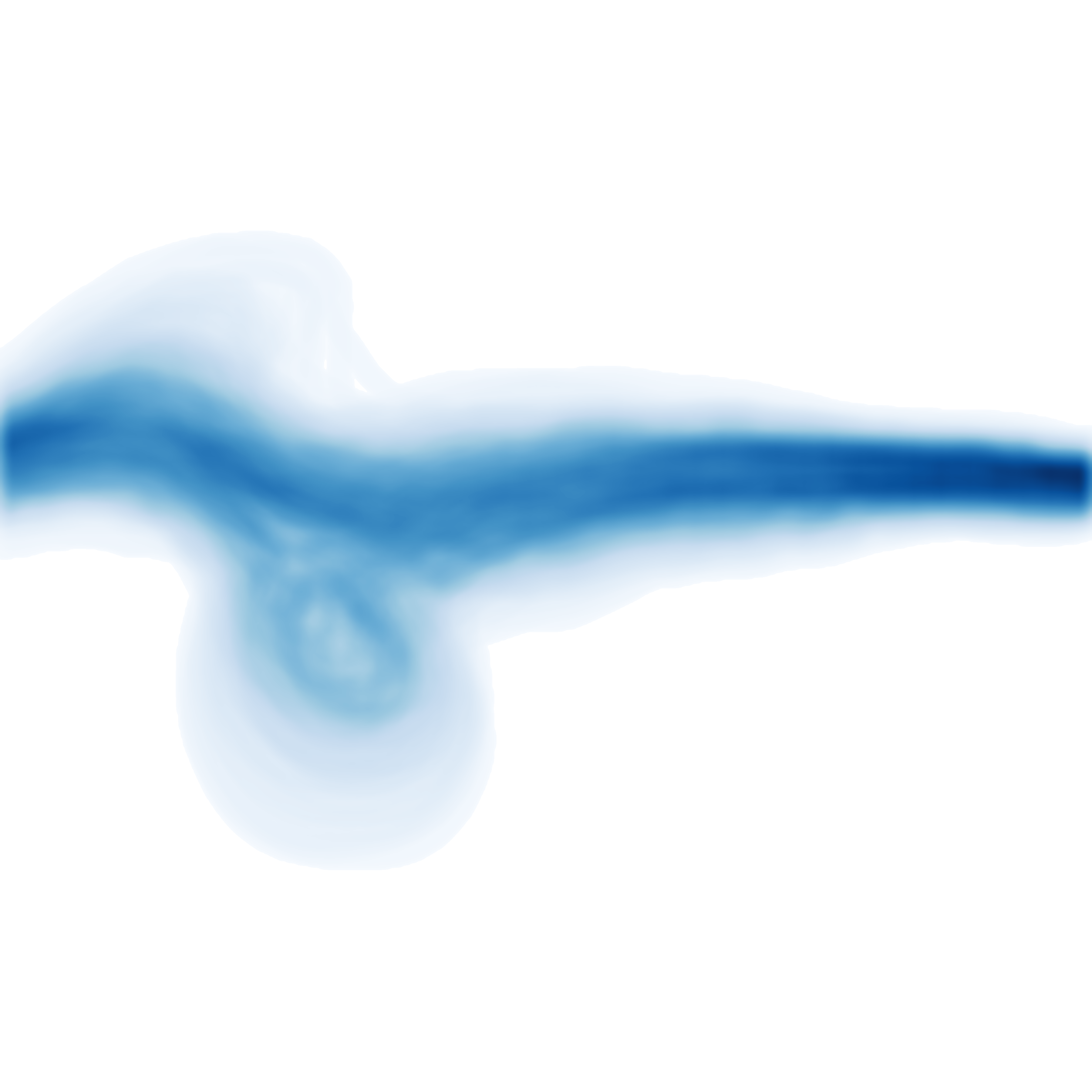}
    }
    \subfloat[Spaghetti plot of member contours]{%
        \includegraphics[width=0.33\textwidth]{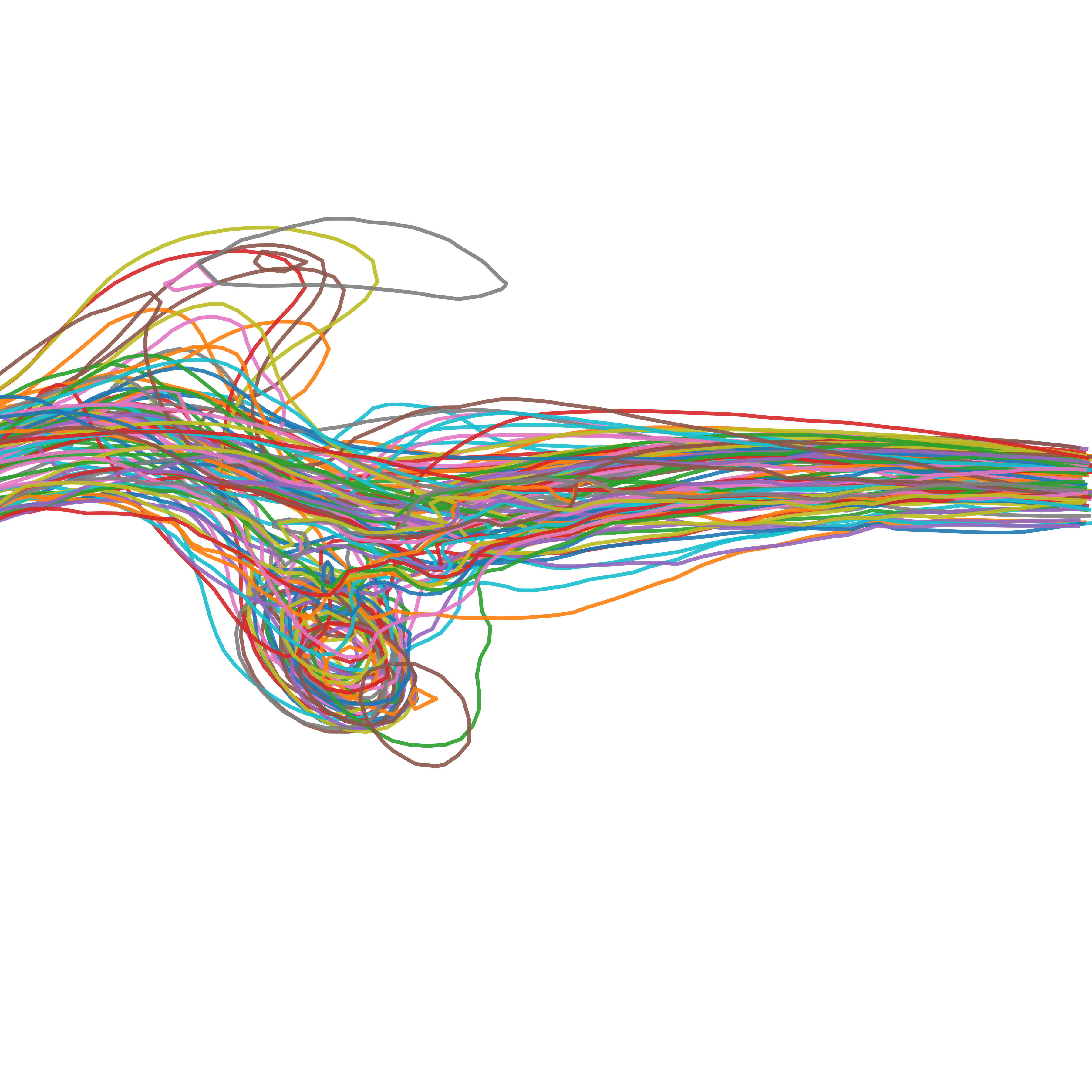}
    }
    \subfloat[Probability density plot with our method]{%
        \includegraphics[width=0.33\textwidth]{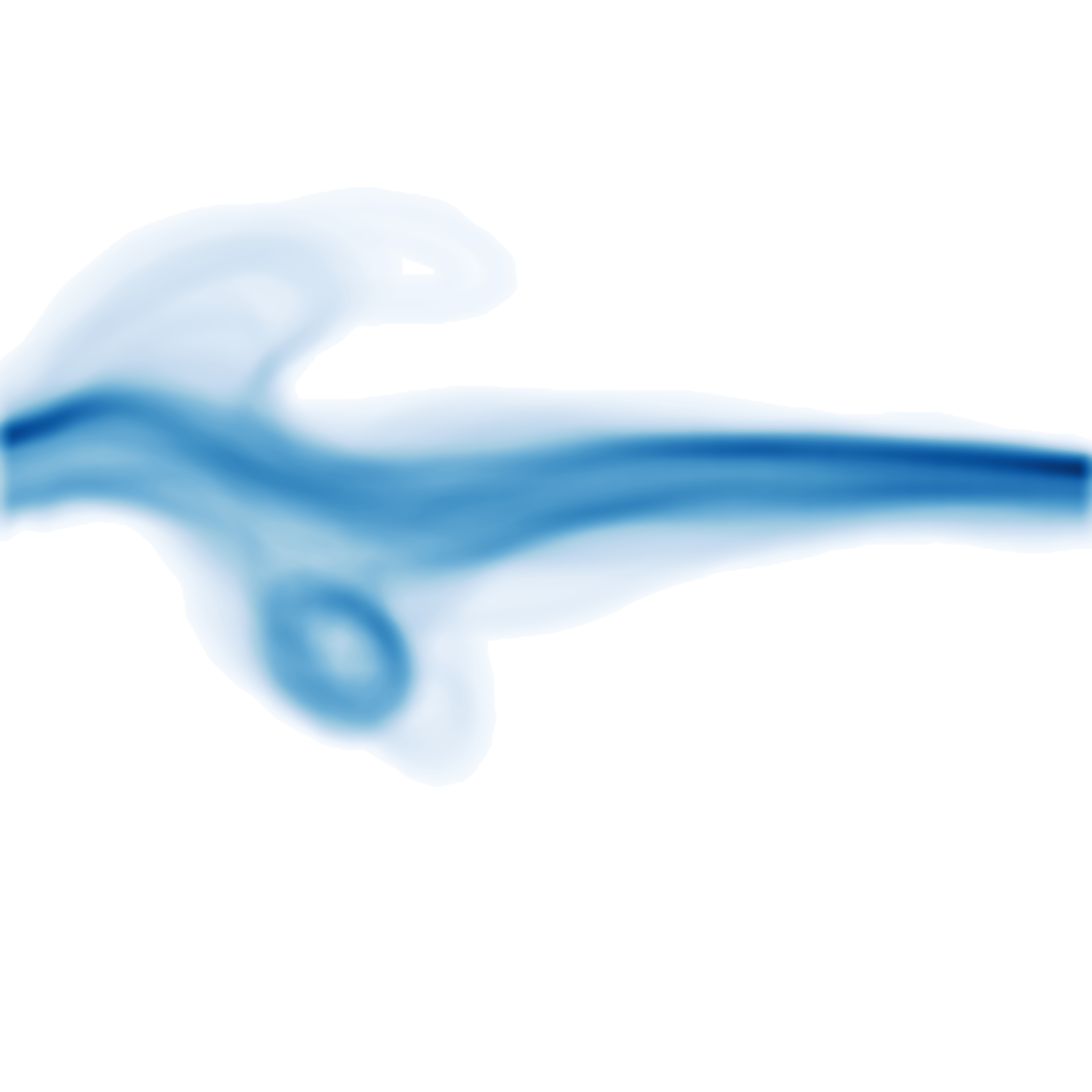}
    }
    \caption{A comparison of density estimation approaches. (a) Probability density plot based on EnConVis, which estimates density at each grid point independently via kernel density estimation; (b) Spaghetti plot of the original data; (c) Probability density plot reconstructed from the VAE-based latent space, which provides more coherent density estimation by jointly capturing local and global features.}
    \label{fig:SupplDensityComparison}
\end{figure*}

\cref{fig:worstCases} shows the top five ensemble member contours that have the largest discrepancies in depth ordering in the evaluation of Section~5.1 of the paper.
For each member, \cref{tab:worstcases} lists the ranks assigned by our MLS depth, the epsilon Inclusion Depth (eID)~\cite{chaves-de-plazaInclusionDepthContour2024}, and the Contour Band Depth (CBD)~\cite{whitakerContourBoxplotsMethod2013}, along with the maximum pairwise rank difference (Max Diff) among the three methods.
As reported in the paper, the overall agreement is strong: our MLS depth is highly correlated with CBD (Pearson $r = 0.933$) and moderately correlated with eID ($r = 0.834$), consistently assigning lower rank values to inner contours and higher rank values to outer contours.
The five cases shown here are the outliers to this general trend, exhibiting the largest rank discrepancies.

As can be seen from~\cref{tab:worstcases}: in all five cases, eID assigns a substantially lower rank (i.e., deeper, more central) than both CBD and our MLS depth, whereas the CBD and MLS ranks are comparatively close to each other.
For example, contour~85 is ranked 22nd by eID but 57th by CBD and 67th by MLS; contour~58 is ranked 48th by eID but 73rd by CBD and 92nd by MLS.
This indicates that eID considers these members to be more centrally located than CBD and our method do.
The likely explanation is that eID uses a local epsilon-neighborhood inclusion criterion that can overestimate the centrality of members with spatially ambiguous contour positions, while CBD and our MLS depth both rely on more global comparisons---combinatorial band containment and probabilistic distributional overlap, respectively---that yield a more consistent assessment of these borderline cases.

More specifically, the three methods differ in their notions of centrality:
\begin{enumerate}
\item \textbf{CBD} relies on pairwise band containment over all triples of contours, which captures a combinatorial notion of how often a contour falls between others.
\item \textbf{eID} approximates CBD by computing an epsilon inclusion relationship, providing an efficient but approximate depth ordering.
\item \textbf{Our MLS depth} measures the probabilistic overlap between the latent distributions of members, incorporating both the structural identity (mean) and encoding uncertainty (variance) of each member.
\end{enumerate}

Despite these conceptual differences, the overall rank agreement between our method and CBD is high (normalized Kendall Tau distance of $0.117$), with the discrepancy to eID ($0.174$) being comparable to that between CBD and eID themselves ($0.160$).
The table confirms this quantitatively: the five largest Max Diff values range from 41 to 45, accounting for less than half of the 95 rank positions, and these worst cases are concentrated in the eID--MLS comparison rather than the CBD--MLS comparison.
Notably, our MLS depth is derived directly from the precomputed similarity matrix $\mathbf{M}$, so the depth computation itself incurs no additional cost beyond what is already required by the pipeline for clustering and density estimation.
\begin{table}[htb]
    \centering
     \caption{Ranks by different data depths for the five members with the largest rank discrepancies. Max Diff indicates the largest rank difference between any two depths for each member.}
     \label{tab:worstcases}
    \begin{tabu}{ccccc}
    \toprule
        Contour ID & eID & CBD & Ours (MLS) & Max Diff  \\
        \midrule
        85 & 22 & 57 & 67 & 45\\
        1 & 38 & 71 & 82 & 44\\
        58 & 48 & 73 & 92 & 44\\
        27 & 35 & 67 & 77 & 42\\
        53 & 31 & 63 & 72 & 41\\
        \bottomrule
    \end{tabu}
\end{table}

\section{Comparison with the variability method}
\label{sec:cvp_comparison}

The contour variational plot (CVP)~\cite{ferstlStreamlineVariabilityPlots2016} uses PCA to reduce the original data space $X$ to a latent space $Z$ and then fits a normal distribution to the data in $Z$.
However, employing this method to generate a probability density plot in the original space introduces two major issues:

\begin{enumerate}
    \item \textbf{Insufficient Nonlinear Feature Extraction:} PCA is inherently a linear dimensionality reduction technique. As a result, the latent space $Z$ constructed by PCA preserves only the linear features of the original data. This limited representation fails to capture the complex, nonlinear structures that are often crucial for accurately describing the data distribution.

    \item \textbf{Irregular Data Distribution in $Z$:} The distribution of data in the latent space $Z$ obtained via PCA is typically irregular. This irregularity makes it challenging to accurately estimate the probability density function (PDF) in $Z$. Consequently, sampling from $Z$ based on an inaccurate density estimate leads to a probability density plot in the original space $X$ that does not faithfully represent the true distribution.
\end{enumerate}

In contrast, our method employs a convolutional VAE~\cite{kingmaAutoencodingVariationalBayes2022} to construct the latent space $Z$.
% Each ensemble member is first converted into a signed distance field (SDF) image via the signed distance transform (Section~3.1.1 of the paper), where the SDF value is negative inside the feature region, positive outside, and zero on the contour boundary.
% This continuous representation preserves boundary geometry and provides richer local variation near the contour compared to binary encoding.
 The VAE encoder maps each ensemble member (as a signed distance field image) to a diagonal Gaussian distribution $\mathcal{N}(\mu_i, \mathrm{diag}(\sigma_i^2))$ in the latent space.
The latent space captures both linear and nonlinear features present in $X$.
Moreover, the $Z$ space generated by the VAE is approximately a multivariate standard Gaussian distribution due to the KL regularization during training, which means that each point in $Z$ has an analytically defined probability density.

To generate density plots, we perform dense posterior sampling: a member $i$ is randomly drawn from the ensemble, and a latent sample $z$ is generated from its posterior $z \sim \mathcal{N}(\mu_i, \mathrm{diag}(\sigma_i^2))$.
Each latent sample is then decoded back into the spatial domain as an SDF image, and its zero-level set---which is inherently a well-defined contour---is extracted using the marching squares algorithm.
This continuity guarantee ensures that every decoded sample produces a geometrically valid contour, preserving the structural integrity of the generated members.
The extracted contours are rasterized onto a high-resolution target grid to accumulate spatial hit counts, and a smooth density field is obtained via FFT-based convolution with a compact 2D smoothing kernel (Section~4.1 of the paper).

\cref{fig:SupplDensityComparison} shows a comparison of three visualizations: \cref{fig:SupplDensityComparison}(a) the probability density plot by CVP using PCA (PCA-CVP), \cref{fig:SupplDensityComparison}(b) the spaghetti plot of the original ensemble members, and \cref{fig:SupplDensityComparison}(c) the probability density plot reconstructed from our VAE-based latent space.
The PCA-CVP density plot (\cref{fig:SupplDensityComparison}(a)) captures the overall horizontal trend of the ensemble but exhibits two notable deficiencies when compared to the actual member distribution shown in the spaghetti plot (\cref{fig:SupplDensityComparison}(b)).
First, the density field appears diffuse and lacks fine-grained detail: contour boundaries are blurred, and the density concentrations do not sharply delineate the regions where ensemble members are most likely to occur.
This is a consequence of the linear nature of PCA, which can only represent directions of maximum variance and fails to capture the nonlinear geometric features that distinguish individual contour shapes (as discussed in Section~4 of the paper).
Second, the PCA-CVP result produces an erroneous over-dispersion in the small cyclonic structure at the bottom of the domain.
As shown in the spaghetti plot (\cref{fig:SupplDensityComparison}(b)), the ensemble members in this region are tightly concentrated, yet the PCA-based density plot spreads the probability mass over a substantially wider area than the actual member distribution warrants.
This artifact arises because the linear latent space of PCA conflates the localized curvature variation of the cyclone with the dominant large-scale variability of the main band, causing the sampled contours to scatter excessively in that region.

In contrast, our VAE-based density plot (\cref{fig:SupplDensityComparison}(c)) closely follows the distribution of the original ensemble members.
The density concentrations align tightly with the contour trajectories visible in the spaghetti plot, preserving both the sharpness of the main horizontal band and the compact structure of the small cyclone at the bottom.
This improved fidelity stems from two properties of our method.
First, the convolutional VAE learns nonlinear latent representations that capture both local geometric details and global structural patterns of contour shapes (Section~3.1 of the paper), enabling it to distinguish the localized cyclonic curvature from the large-scale variability of the main flow.
Second, every decoded sample is a complete, structurally coherent contour reconstructed through $p(x \mid z)$ rather than an independent per-pixel estimate, thereby preserving the global integrity of each generated member (Section~4.1 of the paper).
Together with the well-regularized latent space structure induced by the KL divergence term, these properties ensure that posterior sampling produces geometrically valid contours whose spatial distribution faithfully represents the true ensemble density.

\section{Representative Contour}
\label{sec:representative_contour}
Our MLS-based data depth (Section~3.3 of the paper) ranks ensemble members by their centrality in the latent space $Z$: the deeper a member, the closer it lies to the distributional center.
The deepest \emph{existing} member is the median of the ensemble, and we can further obtain a representative contour as the mean contour.
\subsection{Decoded Central Contour of the Latent Space }
By decoding the center of $Z$---the mean of all posterior means $\bar{\mu} = \frac{1}{N}\sum_{i=1}^{N}\mu_i$---through the VAE decoder, we calculate a \emph{representative contour} that need not coincide with any observed member.
\begin{figure}[htb]
    \centering
    \subfloat[Hippocampus]{\includegraphics[width=0.42\linewidth]{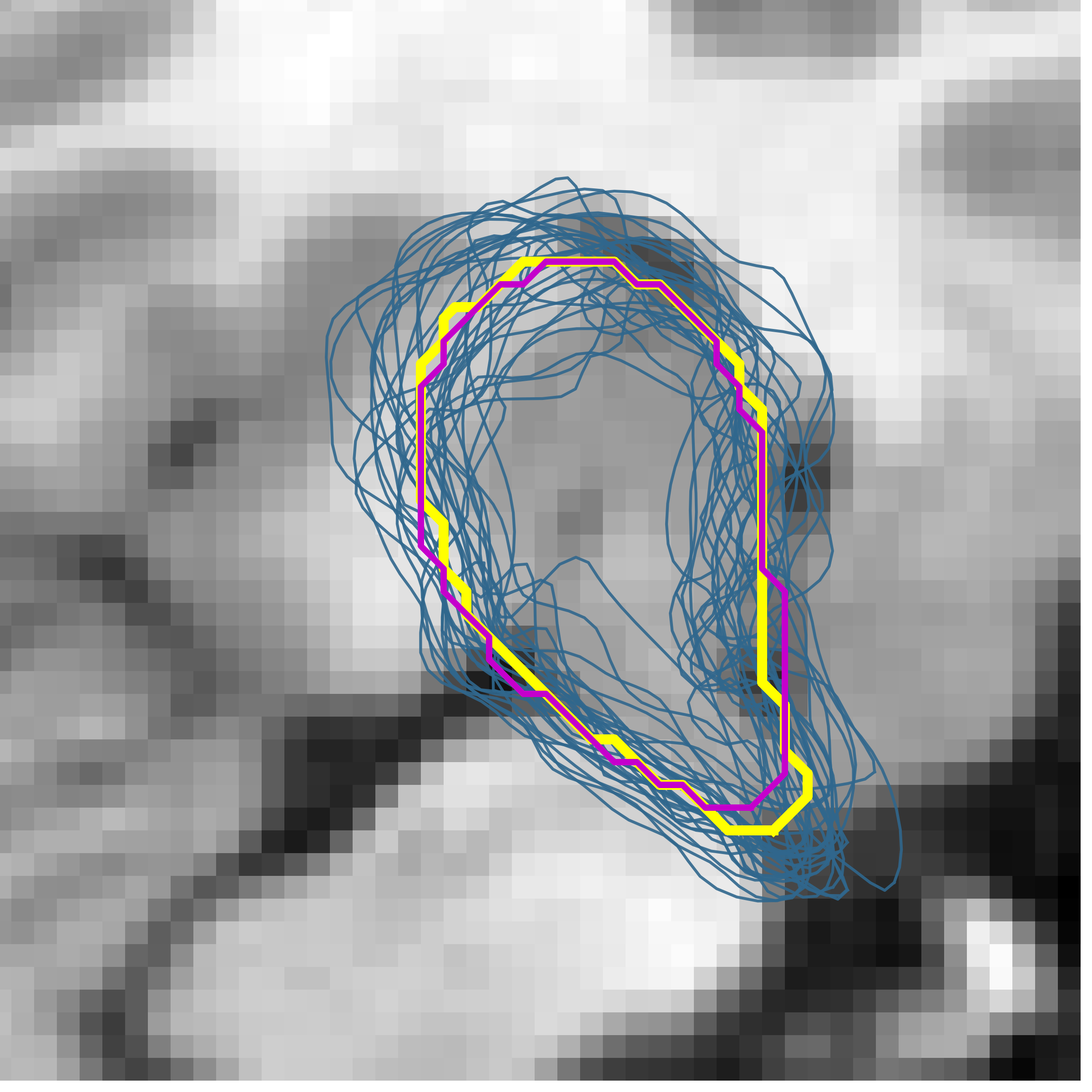}}\hfill
    \subfloat[Third and fourth ventricles]{\includegraphics[width=0.57\linewidth]{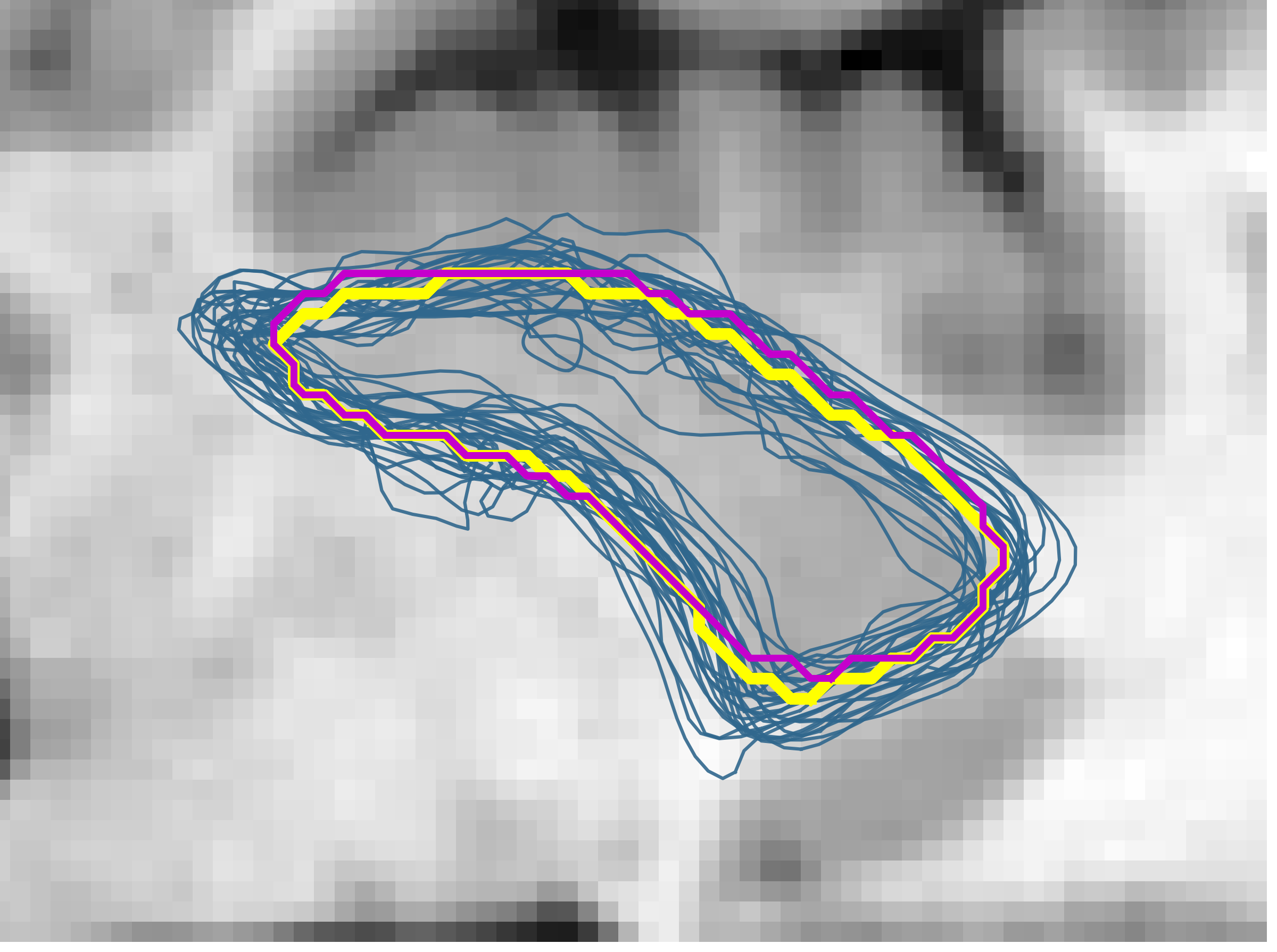}}
        \caption{\rev{Comparison of mean contours for medical image segmentation ensembles from the IXI dataset. The magenta contours are our decoded latent means (representative contours) obtained from the latent space centroid $\bar{\mu}$, and the yellow contours are OT means computed as Wasserstein barycenters. (a) Hippocampus segmentation. (b) Third and fourth ventricle segmentation.}}
    \label{fig:representative_contour_medical}
\end{figure}

This construction is well-motivated for two reasons.
First, as discussed in Section~3.3 of the paper, the probability density of an isotropic Gaussian decays exponentially with the squared distance from its mean, so the log-likelihood of observing any member $z_j$ from a candidate center $z_i$ is proportional to $-\|z_i - z_j\|_2^2$.
Minimizing the average squared distance over the ensemble is therefore equivalent to maximizing the joint log-likelihood, i.e., finding the point of highest probability density.
The point $\bar{\mu}$ is exactly this maximizer, making it the deepest point in the latent space and the natural choice for a representative contour.
Second, the KL regularization during VAE training encourages a smooth, continuous latent space, so that nearby latent points decode to geometrically similar contours.
This smoothness guarantees that the highest-density point in $Z$ decodes to a shape that meaningfully reflects the central tendency of the ensemble, unlike averaging in pixel space where independent per-pixel operations can produce geometrically inconsistent shapes.
Because the decoder maps $\bar{\mu}$ as a single coherent latent point back to the spatial domain as an SDF image whose zero-level set defines the contour, the resulting representative contour preserves global structural integrity.
\begin{figure}[htb]
    \centering
    \includegraphics[trim={0 3cm 0 3cm}, clip, width=\linewidth]{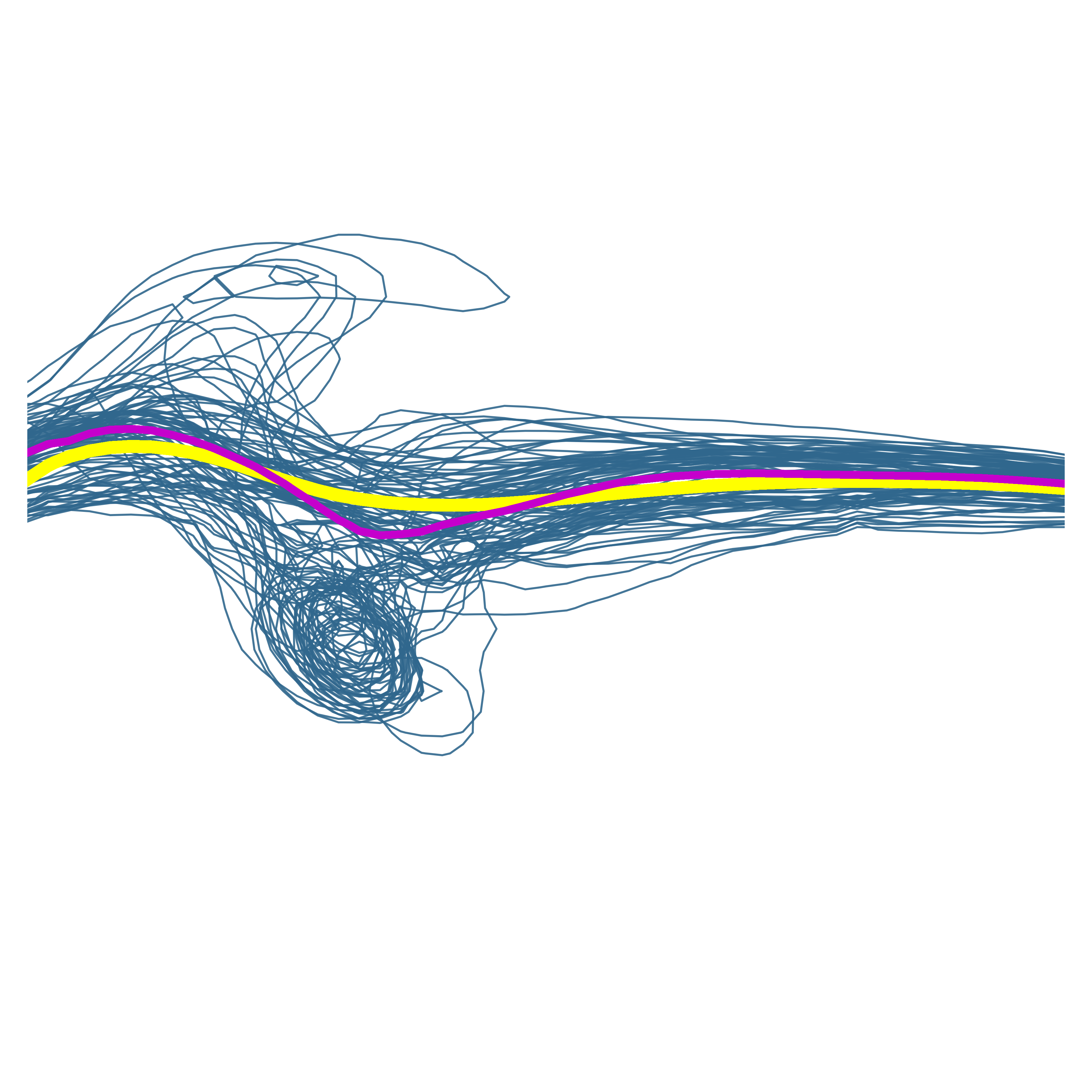}
       \caption{\rev{Comparison of mean contours for the weather simulation ensemble. The magenta contour is our decoded latent mean (representative contour) obtained from the latent space centroid $\bar{\mu}$, and the yellow contour is the OT mean computed as a Wasserstein barycenter.}}

    \label{fig:representative_contour_weather}
\end{figure}
\cref{fig:representative_contour_medical} shows the representative contours for the hippocampus and ventricle segmentation ensembles from the IXI dataset.
The decoded contours delineate boundaries that reflect the consensus across segmentation masks, effectively averaging out inter-subject anatomical variability in the latent space rather than in pixel space.
Despite the distinct morphological characteristics of the two structures, the representative contours faithfully capture the central shape tendency in both cases.
\cref{fig:representative_contour_weather} shows the representative contour for the weather simulation data.
The decoded contour captures the central trajectory of the ensemble while smoothing out member-level fluctuations, providing a compact summary of the dominant flow pattern.

\subsection{Comparison to Mean Contours by Optimal Transport }
\rev{We compare our decoded latent mean with an optimal transport (OT) mean computed as the Wasserstein barycenter~\cite{7053911}. In \cref{fig:representative_contour_medical,fig:representative_contour_weather}, magenta denotes our decoded latent mean (representative contour), whereas yellow denotes the OT mean.}

\rev{The two constructions define a mean with respect to different representations: the Wasserstein barycenter depends on the selected transport cost and regularization, whereas our decoded latent mean represents centrality in the learned probabilistic latent space. The resulting contours are similar in their overall shapes but exhibit visible local differences. Without a ground-truth mean or a task-specific evaluation criterion, neither construction can be considered universally preferable.}

\rev{The decoded latent mean is derived from the same probabilistic representation used for MLS depth, uncertainty-aware clustering, and density estimation, thereby maintaining coherence across these tasks. A systematic comparison of alternative mean-shape definitions under task-specific evaluation criteria is left for future work.}

\section{Synthetic Datasets Generation}
\label{sec:synthetic_data}
Two types of synthetic ensembles are used in our evaluation: a \emph{radial} dataset for the data depth and clustering evaluations (Figs.~9--10 of the paper), and a \emph{cross-shaped} dataset for the multimodal ensemble illustration and the density plot evaluation (Figs.~5 and~12 of the paper).

\subsection{Radial Dataset}
The radial dataset is generated using the function \texttt{main\_shape\_with\_outliers} from the source code of the multimodal contour depth method~\cite{chaves-de-plazaDepthMultimodalContour2024}.
Each ensemble member is a closed contour constructed in polar coordinates.
The generation process is as follows.

\begin{enumerate}
    \item \textbf{Polar coordinate construction.}
    An array of $100$ evenly spaced angles $\theta \in [0, 2\pi]$ is created.
    A uniform base radius of $0.5$ is assigned to every angle, forming a circle centered at the grid center as the template shape.

    \item \textbf{Gaussian-process perturbation.}
    Two sets of smooth perturbations are sampled independently via a Gaussian process (GP) whose covariance kernel operates on both $\sin\theta$ and $\cos\theta$:
    \begin{itemize}
        \item \emph{Normal perturbations} (scale $= 0.003$, length-scale $= 0.9$, with an additive offset of $0.1$) produce gentle shape variations around the template.
        \item \emph{Outlier perturbations} (scale $= 0.009$, length-scale $= 0.04$) produce stronger, higher-frequency deformations.
    \end{itemize}

    \item \textbf{Contamination.}
    Each member is independently assigned to one of two populations with equal probability ($p_{\text{contamination}} = 0.5$).
    For uncontaminated members the normal perturbation is added to the base radius; for contaminated members the outlier perturbation is used instead.
    This creates a bimodal ensemble distribution with known ground-truth labels.

    \item \textbf{Rasterization.}
    The perturbed polar coordinates are converted to Cartesian coordinates and rasterized into binary masks on a $100 \times 100$ grid.
\end{enumerate}

\noindent The following experimental setups are adopted:
\begin{itemize}
    \item \textbf{Depth evaluation} (Section~5.1 of the paper, Fig.~9): $95$ members are generated with a fixed data seed.
    \item \textbf{Clustering evaluation} (Section~5.2 of the paper, Fig.~10): $20$ independently seeded ensembles of $50$ members each are generated, and the clustering accuracy of every method is recorded for each ensemble.
\end{itemize}

\subsection{Cross-Shaped Dataset}
The cross-shaped (X-shaped) dataset is used for the multimodal ensemble illustration (Fig.~5 of the paper) and the density plot evaluation (Section~5.3, Fig.~12).
It is designed to test whether a method can distinguish two spatially overlapping but structurally distinct populations of contours.

Each member is a horizontal curve spanning the full width of a $100 \times 100$ grid.
The ensemble consists of two branches of $50$ members each:
\begin{itemize}
    \item \emph{Branch A} ($\backslash$): the center line follows $y = y_{\text{mid}} + A\cos(\pi x)$, running from the upper-left to the lower-ri  
    \item \emph{Branch B} ($/$): the center line follows $y = y_{\text{mid}} - A\cos(\pi x)$, running from the lower-left to the upper-right.
\end{itemize}
Here $y_{\text{mid}}$ is the vertical midpoint and $A = 30$ pixels controls the opening of the X shape.
The two branches cross at the horizontal center of the grid.

For each member, a smooth random offset is added to the center line by first sampling independent Gaussian noise (standard deviation~$14.0$) at every column and then applying a 1D Gaussian filter (kernel $\sigma = 6$), producing spatially continuous perturbations that mimic the natural variability of ensemble members.
Each member is stored as a 2D scalar field $f(r,c) = r - \text{center}(c)$, so that its level-$0$ contour coincides with the perturbed center line.

This design ensures that the two branches overlap in the central region of the grid.
A pointwise density method treats each spatial location independently and therefore merges the two populations at the crossing, assigning high probability to the empty region between the branches.
In contrast, our VAE-based method encodes entire contours into the latent space and correctly separates the two branches, as demonstrated in Figs.~5 and~12 of the paper.

\section{\rev{Network Configuration and Model Adequacy}}
\label{sec:network_configuration}

\rev{All results in the paper use one common VAE configuration, with three convolutional blocks of channel widths $[64,128,256]$ and latent dimension $k=8$.
We do not tune the architecture for individual datasets.
This section evaluates whether this common configuration is adequate for the 2D contour ensembles considered in the paper and provides a practical procedure for identifying insufficient model capacity.
The goal is not exhaustive neural architecture search or a claim of global optimality.}

\begin{table*}[t]
    \centering
    \caption{\rev{Mean validation NMSE across three training runs for different channel widths with the latent dimension fixed at $k=8$. The configuration used in the paper is shown in bold.}}
    \label{tab:channel_ablation}
    {
    \begin{tabular}{lcccc}
    \toprule
    Dataset &
    $[32,64,128]$ &
    $\mathbf{[64,128,256]}$ &
    $[96,192,384]$ &
    $[128,256,512]$ \\
    \midrule
    Weather
    & $0.001471$
    & $\mathbf{0.001480}$
    & $0.001474$
    & $0.001514$ \\
    Hippocampus
    & $0.009603$
    & $\mathbf{0.011712}$
    & $0.012726$
    & $0.012864$ \\
    Ventricle
    & $0.005755$
    & $\mathbf{0.006339}$
    & $0.007095$
    & $0.007275$ \\
    \bottomrule
    \end{tabular}
    }
\end{table*}

\begin{table*}[t]
    \centering
    \caption{\rev{Mean validation NMSE across three training runs for different latent dimensions with channel widths fixed at $[64,128,256]$. The dimension used in the paper is shown in bold.}}
    \label{tab:latent_ablation}
    {
    \begin{tabular}{lcccccc}
    \toprule
    Dataset &
    $k=2$ &
    $k=4$ &
    $\mathbf{k=8}$ &
    $k=16$ &
    $k=32$ &
    $k=64$ \\
    \midrule
    Weather
    & $0.003141$
    & $0.001886$
    & $\mathbf{0.001480}$
    & $0.001481$
    & $0.001402$
    & $0.001346$ \\
    Hippocampus
    & $0.028012$
    & $0.017494$
    & $\mathbf{0.011712}$
    & $0.011195$
    & $0.009811$
    & $0.009019$ \\
    Ventricle
    & $0.014065$
    & $0.011175$
    & $\mathbf{0.006339}$
    & $0.006483$
    & $0.005723$
    & $0.005621$ \\
    \bottomrule
    \end{tabular}
    }
\end{table*}
\subsection{\rev{Experimental Protocol}}

\rev{We conduct controlled architecture sensitivity experiments on three ensembles with different geometric characteristics: the weather contours, hippocampus segmentations, and third/fourth ventricle segmentations.

For each dataset, a fixed 80/20 split is used for training and validation, and the same held-out members are used for all configurations.
The global mean and variance for SDF normalization are computed from the training members only and then applied unchanged to the validation members.
Each configuration is trained with three initialization seeds ($10$, $42$, and $123$), while the data split and all other optimization settings remain fixed.}

\rev{For deterministic evaluation, a validation member is encoded and reconstructed by decoding its posterior mean $\mu$, rather than a random latent sample.
Let $N_{\mathrm{val}}$ be the number of validation members; let $w$ and $h$ be the width and height of each SDF grid; and let $x_i(u)$ and $\hat{x}_i(u)$ denote the original and reconstructed SDF values at the flattened grid location $u\in\{1,\ldots,wh\}$ for validation member $i$.
Let $\sigma_{\mathrm{train}}^2$ be the variance of the SDF values over all grid locations and training members.
We report the normalized mean squared error}
\begin{equation}
\rev{
\operatorname{NMSE}
=
\frac{
    \frac{1}{N_{\mathrm{val}}wh}
    \sum_{i=1}^{N_{\mathrm{val}}}
    \sum_{u=1}^{wh}
    \left(\hat{x}_i(u)-x_i(u)\right)^2
}{
    \sigma_{\mathrm{train}}^2
}\;.
}
\label{eqn:network_nmse}
\end{equation}
\rev{The reported values are the mean NMSE across the three independent training runs.
In the channel-width sensitivity experiment, the latent dimension is fixed at $k=8$.
In the latent-dimension sensitivity experiment, the channel widths are fixed at $[64,128,256]$.}

\subsection{\rev{Sensitivity to Channel Width}}

\rev{\Cref{tab:channel_ablation} shows that increasing the channel widths beyond the common $[64,128,256]$ configuration does not consistently improve held-out reconstruction.
The weather ensemble is largely insensitive to channel width over the tested range, whereas wider networks increase validation NMSE for the two smaller medical ensembles.
The narrower $[32,64,128]$ network also performs well, confirming that the selected model is not under-capacity with respect to convolutional width.
We retain $[64,128,256]$ in the paper as a single conservative configuration shared by all datasets, rather than selecting a different width after observing each dataset.}

\subsection{\rev{Sensitivity to Latent Dimensionality}}

\rev{\Cref{tab:latent_ablation} shows that very small latent spaces ($k=2$ or $k=4$) produce substantially higher validation errors.
The common $k=8$ configuration reduces NMSE to $0.001480$, $0.011712$, and $0.006339$ for the weather, hippocampus, and ventricle ensembles, respectively.
Increasing the latent dimension from $k=8$ to $k=64$ further reduces these values to $0.001346$, $0.009019$, and $0.005621$, respectively; however, because $k=8$ already reaches a low-error range, the corresponding absolute NMSE reductions are limited.
We therefore do not interpret $k=8$ as the exact intrinsic dimensionality of the feature space or as a universally reconstruction-optimal value.
Estimating intrinsic dimensionality reliably from these relatively small ensembles is a separate problem; here, $k$ is an architectural bottleneck whose adequacy is evaluated empirically.
We use $k=8$ as a compact common representation that already achieves low held-out NMSE and supports all downstream analyses in the paper, while avoiding dataset-specific tuning and the additional cost of larger latent spaces.}

\subsection{\rev{Recommendations for New Datasets}}

\rev{For a new ensemble, we recommend starting from the common compact configuration and reserving a small set of members for validation.
A high validation NMSE indicates insufficient representational capacity and motivates increasing the channel width, network depth, or latent dimension.
Because a pixelwise SDF error alone may not expose failures at the zero level set, the reconstructed and posterior-sampled SDFs should also be checked for a valid zero crossing and for severe artifacts such as missing, fragmented, or out-of-domain contours.
A configuration is considered adequate when it attains low held-out NMSE and produces valid decoded contours for the intended visualization tasks; it need not minimize reconstruction error globally.}

\rev{The sensitivity experiments above cover the 2D contour ensembles used in this work.
Substantially more complex or higher-resolution inputs may require deeper or wider encoders and decoders, or a larger latent space.
The same validation protocol provides a practical way to determine whether such additional capacity is beneficial.
A systematic evaluation on substantially more complex 2D data and large 3D ensembles remains future work.}

\section{\rev{Measured Running Time}}
\label{sec:measured_runtime}

\begin{table}[t]
    \centering
    \caption{\rev{Measured stage-wise running times in seconds. (a) Timings of the proposed method on all five datasets. (b) Task-matched comparisons on Weather: MLS-AHC is compared with PCA--AHC and CD-clustering, MLS depth with eID, and our density construction with EnConVis.}}
    \label{tab:measured_runtime}
    \scriptsize
    \setlength{\tabcolsep}{3.5pt}
    {
    \textbf{(a) Timings of the our method across datasets (s) }\par\vspace{0.25em}
    \begin{tabular}{lrrrrrrr}
    \toprule
    Dataset & SDF & Train & Enc. & Matrix & Cluster & Depth & Density \\
    \midrule
    Weather
    & 0.0221 & 57.3730 & 0.0042 & 0.0183 & 0.0548 & 0.0004 & 4.6080 \\
    Hippocampus
    & 0.0049 & 13.9274 & 0.0005 & 0.0019 & 0.0017 & 0.0001 & 6.8324 \\
    Ventricle
    & 0.0058 & 15.7970 & 0.0006 & 0.0020 & 0.0017 & 0.0001 & 9.0520 \\
    Synthetic X
    & 0.0109 & 54.3923 & 0.0027 & 0.0203 & 0.0635 & 0.0004 & 5.7596 \\
    ScalarFlow
    & 0.0175 & 65.9186 & 0.0030 & 0.0218 & 0.0717 & 0.0004 & 1.7258 \\
    \bottomrule
    \end{tabular}
    \par\vspace{0.75em}

    \textbf{(b) Task-matched comparison on Weather}\par\vspace{0.25em}
    \begin{tabular}{llrlr}
    \toprule
    Task & Our method & Time (s) & Baseline method & Time (s) \\
    \midrule
    Clustering & MLS-AHC & 0.0548 & PCA--AHC & 0.0709 \\
    Clustering & MLS-AHC & 0.0548 & CD-clustering & 0.1056 \\
    Depth & MLS depth & 0.0004 & eID & 0.1071 \\
    Density & Ours & 4.6080 & EnConVis & 0.4746 \\
    \bottomrule
    \end{tabular}
    }
    \vspace{0.25em}

    \parbox{0.98\linewidth}{\scriptsize\rev{The PCA--AHC measurement includes contour extraction, contour resampling, PCA projection, and AHC; the CD-clustering measurement includes its required mask construction and clustering. Encoding and MLS-matrix construction for our method are reported separately in (a) because they are shared by clustering and depth. To avoid duplicating nearly identical measurements, the larger observed eID and EnConVis times from the two baseline workflows are reported.}}
\end{table}
\rev{All timings were measured on the workstation described in the paper---a machine with a 3.4 GHz Intel i7 CPU, 32 GB main memory, and an NVIDIA GeForce RTX 4080 Super GPU with 16 GB graphics memory, using PyTorch~2.6 and CUDA~12.4.
Each VAE was trained from scratch for 3,000 epochs with the configuration used in the corresponding experiment.
GPU operations were synchronized at the boundaries of every timed stage.
For a consistent comparison, we computed the SDFs for all members, encoded the ensemble once, constructed one shared MLS matrix, performed clustering once, computed only the global depth, and generated one global density map from 3,000 posterior samples.
}

\rev{\Cref{tab:measured_runtime}(b) reports task-level running-time comparisons on the Weather dataset.
For Weather, MLS-AHC requires 0.0548~s, compared with 0.0709~s for PCA--AHC~\cite{ferstlStreamlineVariabilityPlots2016} and 0.1056~s for CD-clustering~\cite{chaves-de-plazaDepthMultimodalContour2024}.
MLS depth requires 0.0004~s, compared with 0.1071~s for eID~\cite{chaves-de-plazaInclusionDepthContour2024}.
In contrast, our density construction requires 4.6080~s, whereas EnConVis~\cite{zhangEnConVisUnifiedFramework2023} requires 0.4746~s.
Thus, after the shared representation is available, our clustering and depth stages are fast, but our generative density construction is slower than EnConVis.
When the one-time VAE training cost is included, our method is not faster for a single analysis: training alone requires 57.3730~s on Weather and ranges from 13.9274~s to 65.9186~s across the five datasets.}

\rev{The trained model, encoded member distributions, and complete pairwise MLS matrix are reusable.
If users explore several values of the cluster number, repeat clustering with different settings, or compute depth separately within multiple resulting clusters, the VAE does not need to be retrained, the ensemble does not need to be encoded again, and the full MLS matrix does not need to be reconstructed.
Each new clustering operates on the cached MLS matrix, and a within-cluster depth analysis uses the corresponding submatrix and row averages.
For example, on Weather, one-time encoding and MLS-matrix construction require 0.0042~s and 0.0183~s, respectively, while one clustering and one global-depth computation require only 0.0548~s and 0.0004~s.
Consequently, repeated clustering and within-cluster depth analyses incur only the downstream costs reported in the table; training, encoding, and construction of the complete MLS matrix remain fixed preprocessing costs and are not repeated.
This reuse should not be interpreted as an overall runtime advantage over the conventional methods.
}

% \subsection{\texttt{pdfendlink} error}

% Occasionally (for some \LaTeX\ distributions) this hyper-linked bib\TeX\ style may lead to \textbf{compilation errors} (\texttt{pdfendlink ended up in different nesting level ...}) if a reference entry is broken across two pages (due to a bug in \verb|hyperref|).
% In this case, make sure you have the latest version of the \verb|hyperref| package (i.e.\ update your \LaTeX\ installation/packages) or, alternatively, revert back to \verb|\bibliographystyle{abbrv-doi}| (at the expense of removing hyperlinks from the bibliography) and try \verb|\bibliographystyle{abbrv-doi-hyperref}| again after some more editing.

\end{document}